\documentclass[useAMS,usenatbib]{mn2e}
\usepackage{graphicx,color}
\usepackage[colorlinks=true,citecolor=blue,linkcolor=red,urlcolor=blue]{hyperref}
\usepackage{dcolumn}
\usepackage{url}
\usepackage{times}
\newcommand{\Msun}{\ensuremath{M_{\odot}}}
\newcommand{\Zsun}{\ensuremath{Z_{\odot}}}
\newcommand{\Ha}{H$\alpha$ }

\newcommand{\flux}{erg\,cm$^{-2}$\,s$^{-1}$}
\newcommand{\lum}{erg\,s$^{-1}$\,}

\newcommand{\tim}{$\times\,$}

\newcommand{\s}{$\sim\, $}
\newcommand{\de}{$\,^{\circ}$}

\title[Cavities in ZwCl~2701]{AGN driven perturbations in the Intra-cluster medium of cool core cluster ZwCl~2701}
\author[Vagshette et.al.]{Nilkanth D. Vagshette$^{1}$\thanks{E-mail: nilkanth@prl.res.in}, Satish S. Sonkamble$^{2}$\thanks{E-mail: satish04apr@gmail.com}, Sachindra Naik$^{1}$\thanks{E-mail: snaik@prl.res.in}, Madhav. K. Patil$^{2}$\thanks{E-mail: patil@associates.iucaa.in} \\ \\
$^{1}$ Physical Research Laboratory, Navrangpura, Ahmedabad - 380 009, India\\
$^{2}$School of Physical Sciences, Swami Ramanand Teerth Marathwada University, Nanded - 431 606, India.\\
}
\begin{document}
\pagerange{\pageref{firstpage}--\pageref{lastpage}} \pubyear{2016}
\maketitle
\label{firstpage}

%=================================================================================================
\begin{abstract}
We present the results obtained from a total of 123 ks X-ray (\textit{Chandra}) and 8 hrs of 1.4 GHz radio 
(Giant Metrewave Radio Telescope - GMRT) observations of the cool core cluster ZwCl~2701 ($z$ = 0.214). These observations of ZwCl~2701 showed the presence of an extensive pair of ellipsoidal cavities along the East and 
West directions within the central region $<$ 20 kpc. Detection of bright rims around the cavities suggested 
that the radio lobes displaced X-ray emitting hot gas forming shell-like structures. The total cavity power 
(mechanical power) that directly heated the surrounding gas and cooling luminosity of the cluster were estimated 
to be \s 2.27 \tim $10^{45}$ \lum and $3.5 \times\, 10^{44}\,$ \lum, respectively. Comparable values of cavity 
power and cooling luminosity of ZwCL~2701 suggested that the mechanical power of the AGN outburst is large enough 
to balance the radiative cooling in the system. The star formation rate derived from the \Ha luminosity was 
found to be \s 0.60 \Msun yr$^{-1}$ which is about three orders of magnitude lower than the cooling rate of 
\s 196 \Msun yr$^{-1}$. Detection of the floor in entropy profile of ZwCl~2701 suggested the presence of an 
alternative heating mechanism at the centre of the cluster. Lower value of the ratio ($\sim$10$^{-2}$) between 
black hole mass accretion rate and Eddington mass accretion rate suggested that launching of jet from the super 
massive black hole is efficient in ZwCl~2701. However, higher value of ratio ($\sim$10$^{3}$) between black 
hole mass accretion rate and Bondi accretion rate indicated that the accretion rate required to create cavities 
is well above the Bondi accretion rate.

\end{abstract}
\begin{keywords}
galaxies:active-galaxies:general-galaxies:clusters:individual:ZwCl~2701-Intracluster medium-X-rays:galaxies:clusters
\end{keywords}
%==================================================================================================

\section{Introduction}
Active galactic nuclei (AGN) feedback plays an important role in the evolution of non-thermal 
radio jets and lobes (radio bubbles) ejected from the central supermassive black hole (SMBH). 
These jets and lobes, filled with relativistic plasma, interact with the surrounding environment 
releasing a large amount of energy. Such examples are seen in $\sim$70\% relaxed, cool-core galaxy 
clusters (\citealt{1990AJ.....99...14B,2006MNRAS.373..959D,2007MNRAS.379..894B,2009A&A...501..835M}). 
The brightest cluster galaxy (BCG) at the core of the cool core clusters is found to be radio 
loud galaxy (\citealt{2015aska.confE..76G}) where jets from the AGN in the galaxy extend outwards 
in a bipolar flow and inflate the lobes. These lobes push the X-ray emitting hot gas in the 
Intra-cluster Medium (ICM) and 
create X-ray deficiency regions called as ``cavities" \citep{2006ApJ...652..216R}. The availability 
of high angular resolution X-ray imaging detectors onboard \textit{Chandra} observatory made it 
possible to detect such cavities in several of the clusters (\citealt{2006ApJ...652..216R, 2008ApJ...686..859B, 2010MNRAS.404..180D, 2011MNRAS.416.2916O, 2009ApJ...705..624D, 2013Ap&SS.345..183P, 2012AdAst2012E...6G}). It is believed that these cavities are filled with non-thermal radio lobes consisting of relativistic particles and magnetic field. This has been supported by the observations 
of radio lobes that spatially match with the X-ray cavities in many clusters (\citealt{2012AdAst2012E...6G, 2007ARA&A..45..117M}). Independent studies of Perseus cluster 
by several researcher have demonstrated that the AGN in the cluster has sufficient power to 
offset the cooling flow by inflating cavities inducing weak shocks and sound waves (\citealt{2006MNRAS.366..417F, 2005ApJ...625L...9N, 2005ApJ...635..894F, 2007ApJ...665.1057F, 2007MNRAS.381.1381S, 2003MNRAS.344L..43F, 2006MNRAS.366..417F, 2004ApJ...607..800B, 2006MNRAS.373..959D, 2008MNRAS.385..757D, 2006ApJ...652..216R, 2007ARA&A..45..117M}). A large number of studies of 
cavities in cluster galaxies also provided useful information on the AGN feedback mechanism (\citealt{2004ApJ...607..800B, 2008ApJ...686..859B, 2005MNRAS.364.1343D, 2006MNRAS.373..959D, 2008MNRAS.385..757D, 2007AAS...210.3407N, 2010ApJ...720.1066C, 2010MNRAS.404..180D, 2010ApJ...712..883D, 2011ApJ...735...11O, 2012MNRAS.421.1360H, 2012MNRAS.421..808P, 2013Ap&SS.345..183P, 2015Ap&SS.359...61S}).
 
Positive gradient in the azimuthally averaged radial temperature profiles and short cooling times 
in the central region are the characteristic features of the cool core clusters and have been 
confirmed through numerous studies employing high resolution \textit{Chandra} and \textit{XMM-Newton}
observations (\citealt{1994PASJ...46L..55F,   2004A&A...413..415K, 2006MNRAS.372.1496S, 2012MNRAS.421.1360H}). Temperature of the ICM has been estimated to be $T_e = 10^7\,-\,10^8$ K 
whereas the particle number density was found to be in the range of $10^{-2}$ cm$^{-3}$ at core 
of the clusters to $10^{-4}$ cm$^{-3}$ in the outer regions (\citealt{2007PhR...443....1M}). The 
ICM mainly consists of highly ionized hydrogen and helium and traces of ionized heavier elements 
(about one third of the solar abundance) that gradually increase to the solar value at the centre 
of the cluster. The higher gas density at the centre of the cluster corresponds to shorter
cooling time and large flow of cooling material known as ``cooling flow'' 
(\citealt{1986RvMP...58....1S, 1984Natur.310..733F, 1994ARA&A..32..277F}). However, large 
number of \textit{Chandra} and \textit{XMM-Newton} observations showed the presence of significantly 
less amount of cooled material against that expected from the standard cooling flow model (\citealt{2001A&A...365L.181B, 2001A&A...365L..87T,2001A&A...365L.104P, 2003ApJ...590..207P, 2006PhR...427....1P, 2007ARA&A..45..117M}). 

Radio sources associated with the central AGN release large amount of kinetic energy in the 
form of jets and lobes (\citealt{1984RvMP...56..255B,2000ApJ...534L.135M,2005ApJ...635..894F,2005ApJ...628..629N}). 
If the jets/lobes are powerful enough to get coupled effectively with the surrounding 
ICM of the host galaxy, it will prevent the cooling of the ICM and hence the star formation \citep{2004MNRAS.355..862D,2006ApJ...652..216R,2010ApJ...720.1066C}. Thus, the AGN feedback 
mechanism plays an important role in regulating the star formation in the cores of such 
clusters. In such cases, it is believed that the radio emission from the AGN and star 
formation in the core of a cluster may be related with the ICM-specific entropy distribution (\citealt{2006ApJ...643..730D,2008ApJ...683L.107C}). Being the fundamental quantity of the 
ICM, entropy measurement provides important information regarding the cooling and heating 
processes that are involved in the cores of the cluster (\citealt{2002ApJ...576..601V, 2005MNRAS.364..909V, 2008ApJ...683L.107C}). It holds the record of accretion history and 
influence of non-gravitational effects on the ICM (\citealt{2010A&A...511A..85P}). \cite{2008ApJ...683L.107C} studied the intra-cluster entropy profiles for a sample of 222 
galaxy clusters and showed that \Ha and radio emission are prominent when the entropy ``floor'' 
in the core region is less then a 30 keV cm$^{2}$. 

Apart from AGN feedback, there are other possible sources of ICM heating as
suggested by several authors, which include merger, conduction, supernovae, cosmic ray heating
etc. It is understood that several clusters are being formed because of merging of smaller 
mass concentrations through gravitational infall of matter. During merger process large 
amount of energy is carried by the gas which is being dissipated in the form of shocks and turbulence 
(\citealt{2007PhR...443....1M,2006PhR...427....1P}). The energy dissipated through shock
and turbulence enhances the temperature of the ICM. Due to availability of high resolution 
imaging instruments onboard {\it Chandra} and {\it XMM-Newton} observatories, it is 
possible to identify these merger features in the form of cold fronts, shock fronts, sloshing, 
etc. The cold fronts are seen in the ICM when the gas temperature outside the edge is much 
higher than the inside. These cold fronts are known to be formed because of the ram pressure 
acting on the core of one of the cluster moving into the other cluster at high velocity. 
For a detailed description on merging features in clusters e.g. cold fronts, shocks, sloshing 
etc. refer to \citealt{2007PhR...443....1M} and references therein. The outer atmosphere 
(beyond cooling radius) of cluster represents the reservoir of huge amount of thermal 
energy. The conduction of heat from the outer hot ICM into the cooler central region also 
act as a source of heating (\citealt{2004MNRAS.347.1130V,2003ApJ...582..162Z}). 
Supernova (SN) explosions also act as a heating source and can cause delay in the 
formation of cooling flow (\citealt{2004A&A...425L..21D}). Another mechanism
of ICM heating is known to be due to the cosmic rays (\citealt{2004A&A...413..441C,2004PhRvL..92s1301T}). 
Relativistic electrons form a minihalo in the core of the cluster and produce hard X-ray photons 
due to inverse Compton scattering. These relativistic electrons and hard X-ray photons heat the 
gas in the ICM (\citealt{2006PhR...427....1P}).

ZwCl~2701 is a known relaxed galaxy cluster (\citealt{2015A&A...580A..97C,2015A&A...581A..23K}) 
positioned at RA = 09:52:49.20, DEC = +51:53:05.19 and located at a redshift of 0.214. The 
presence of X-ray cavities in this system have been reported by \cite{2006ApJ...652..216R} in 
a broad study of a sample of 31 groups and cluster galaxies. \cite{2008ApJ...687..986D} also 
carried out similar kind of study and estimated masses of black holes located in the center of 
the brightest galaxies in groups and clusters including ZwCl~2701. The measure of radio flux 
(or luminosity) and their correlation with AGN-power (cavity power) of the clusters (including 
ZwCl~2701) and group galaxies was done by \cite{2008ApJ...686..859B}. Though this object has 
been studied along with several other clusters, a systematic study with an emphasis on the 
comparison of the X-ray and radio properties was not available in the literature. This paper 
presents a systematic study of imaging and spectroscopic properties of ZwCl~2701 using deep 
\textit{Chandra} X-ray data and 1.4\,GHz {\it GMRT} observations. The thermodynamical parameters 
of the ICM selected from different regions of interest from ZwCl~2701 are presented in this paper. 
The cavity detection and their energy content along with balance between the cavity power 
($P_{cav}$) and the radiative loss of the X-ray emission due to its cooling within the cooling 
region ($L_{ICM}$) have also been examined in this paper. Similar balance has been reported in 
several other cooling flow clusters (for e.g., \citealt{2004ApJ...607..800B, 1995MNRAS.276..663B, 1997ApJ...484..602T, 2001ApJ...551..131C,2001ApJ...549..832S,2012MNRAS.424.2971O,2012MNRAS.421.1360H,2013ApJ...777..163H,2014MNRAS.444..651M}). With an objective to investigate signatures of the interactions between the radio jets/lobes and 
the surrounding ICM, we tried to find association between the X-ray cavities and the radio emission 
maps derived from the analysis of 1.4 GHz {\it GMRT} data.
 
The structure of this paper is as follows: Section 2 describes the \textit{Chandra} and 
\textit{GMRT} observations and their reduction strategies. In Section~3, we report the results obtained 
from the imaging analysis of X-ray emission and estimation of the thermodynamical parameters of the X-ray 
emitting gas from different regions of interest through X-ray spectroscopy. This section also presents 
the detection and  estimation of the energetics associated with X-ray cavities in this cluster. 
Section~4 \& 5 provides brief discussion on the results derived from this study  and conclusions, 
respectively. Throughout this paper we assume the $\Lambda$CDM cosmology with $H_0$ = 70 km s$^{-1}$ 
Mpc$^{-1}$, $\Omega_M$ = 0.3, and $\Omega_{\Lambda}$ = 0.7 at a redshift of $z$ = 0.214 for the 
system ZwCl~2701. This cosmology corresponds to a scale of 1\arcsec\,= 3.5 kpc and the luminosity 
distance of the cluster $\sim$ 1022 Mpc. 

%============================================
\begin{figure*}
\vbox{
\centering
\includegraphics[height=6cm, width=14cm]{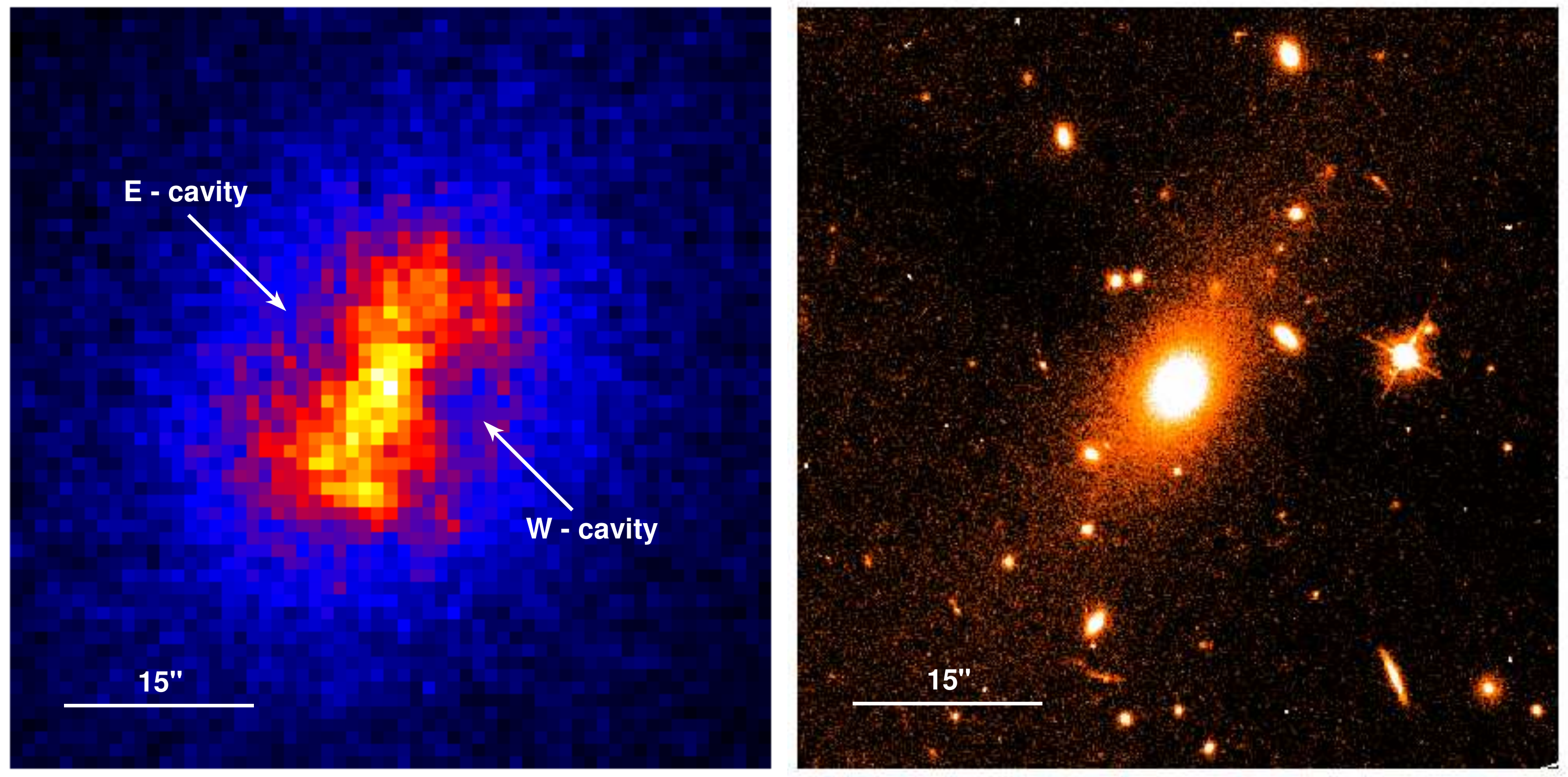}
}
\caption{Background subtracted exposure corrected $1'\times1'$ \textit{Chandra} image 
of ZwCl~2701 and wide band 606~nm \textit{HST} $1' \times 1'$ image are shown in the left 
and right panels, respectively. The arrows in the {\it Chandra} image indicate the presence of X-ray deficiency regions in the cluster.}
\label{fig1}
\end{figure*}
%============================================

\section{Observations and data reduction}
\subsection{Chandra X-ray data}
ZwCl~2701 was observed with \textit{Chandra} on 03 February 2011 in imaging mode for $\sim$96 ks 
(ObsID 12903) and on 04 November 2001 for 27 ks (ObsID 3195) with the source focused on the 
back-illuminated ACIS-S3 chip. Level-1 X-ray data sets available in the public domain of 
\textit{Chandra X-ray Centre}\footnote{\color{blue} {{http://cda.harvard.edu/chaser/}}} have 
been used for the present study. The data sets were reprocessed following the standard data 
reduction routines as described in \textit{Chandra Interactive Analysis of Observations} (CIAO-4.6)\footnote{\color{blue}{{{http://cxc.harvard.edu/ciao/threads/index.html}}}}. Latest 
calibration files (CALDB V~4.6.2) were used in our data analysis. Bad pixels and other artefacts 
were identified and removed from the level~1 data sets by running the script \textit{chandra$\_$repro}. 
As ZwCl~2701 is positioned on chip S3, where the gas distribution of the cluster may cover the entire chip, the background light curve was extracted from the back-illuminated ACIS-S1 chip. The script \textit{lc$\_$sigma$\_$clip} within \textsc{Sherpa} was used for the removal of the segments of high background flaring events (with 3$\sigma$ clipping threshold). Light curve obtained from the dataset 
with ObsID 3195 showed a few flaring episodes with count rate exceeding the mean count rate during the observation. Removal of such flaring segments from the data reduced the effective exposure of the observation to 24.4 ks. However, there were no flares in the light curve obtained from the observation with ObsID 12903.

We used ACIS-S blank-sky observations provided by the CXC for sky background subtraction from the 
science data. CIAO task \textit{$acis\_bkgrnd\_lookup$} was used to identify the blank-sky data corresponding to ZwCl~2701 observations and the background count rates in 9 - 12 keV band were 
adjusted so as to match those in the source image (\citealt{2002astro.ph..5333M}, \citealt{2009ApJ...705..624D}). As ZwCl~2701 observations were performed in Very Faint (VFAINT) 
telemetry format, VFAINT background screening was applied to the blank sky file after including 
events with ``status=0". Exposure maps and exposure corrected images of ZwCl~2701 in 0.3 - 3 keV 
energy band were created using standard CIAO \textsc{fluximage} script. Contaminating point sources 
across the field of view detected through the CIAO script \textit{wavdetect} (\citealt{2013NewA...21....1V}) were excluded from further analysis. The source spectra, background spectra, response and effective area files were created by using \textit{specextract} task of CIAO. Spectral fitting was performed by using XSPEC-12.8.1 package.

\subsection{GMRT radio data}
ZwCl~2701 cluster was observed with the Giant Meterwave Radio Telescope (\textit{GMRT}), Pune, India at frequency of 1.4 GHz (L-band) with 32-MHz band width divided in 512 channels (Project Code 26\_064). The observations were carried out in two observing runs, on 7 September 2014 and 13 September 2014, for \s 4 hours duration in each run. The observations were performed using \textit{GMRT} hardware back-end in upper and lower side bands (USB and LSB). The data were recorded in the LL and RR polarizations with an integration time per visibility of 16 sec during both the observations. The standard point like radio sources 3C~147 and 0834+555 were used as flux density and phase calibrator, respectively. 

The data calibration and reduction was carried out using the NRAO Astronomical Image Processing System (AIPS) package (Version 31DEC2013). We followed the standard data reduction procedure of \textit{GMRT} observations as described in \cite{2011ApJ...732...95G, 2014ApJ...795...73G}. The normal flagging was done by removing non-working antennae, bad baselines, channels and time ranges. Later careful editing was carried out to identify and remove those visibilities affected by the radio frequency interference (RFI). The visibilities were then calibrated for the bandpass response of the antennas using the flux calibrator 3C~147.The source and calibrator flux densities were set to the best \textit{VLA} values suggested by \cite{Per1999} extension to the \cite{1977A&A....61...99B} scaled at 1.4 GHz using SETJY task of AIPS.  Few noisy channels were seen at the end of each band, which were then discarded from the further analysis. Several cycles of self-calibration (using IMAGR and CALIB task) were performed in order to get the image. The residual phase variations from the data sets were also removed so as to obtain the good quality image of the source ZwCl~2701. 

The measured values of noise on the USB and LSB were found to be \s 26.7 $\mu$Jy/beam, which were corrected from \textit{GMRT} primary beam response\footnote{\color{blue}http://www.ncra.tifr.res.in:8081/\%7Engk/primarybeam/beam.html} using PBCOR task of AIPS. The resultant flux image was obtained after combining the corrected USB and LSB data sets. The  ZwCl~2701 cluster has been detected in the flux image at position of RA=09:52:49.14 and DEC=+51:53:06.2. The flux density of ZwCl~2701 cluster was derived by fitting the Gaussian model to the source using JMFIT task. The restoring beam image was obtained with the Briggs ``robustness'' parameter set to -3 (ROBUST = -3) in IMAGR task for uniform weighting. Final image of the cluster was obtained at the beam with FWHM \s  2.82\arcsec \tim 1.27\arcsec\, resolution with position angle of 84\de. The 1$\sigma$ rms noise level of 26.7 $\mu$Jy/beam was achieved at the center of the source in the final image.

\section{Results}
\subsection{X-ray Imaging}
Background subtracted and exposure corrected 0.3 - 3 keV \textit{Chandra} image of ZwCl~2701 is shown 
in Figure~\ref{fig1} (\textit{left panel}) whereas the image in the \textit{right panel} delineates 
\textit{HST} 606 $nm$ image of the cluster. To examine the presence of any offset between optical 
and X-ray centers of the cluster, we identified the optical and X-ray peaks in corresponding images and
found that both the peaks are well aligned with a maximum offset of $\leq$0.5\arcsec. Both of these images 
are well aligned along the North-West and South-East directions. Several extended sources surrounding the central brightest object are clearly evident within  1$'$ $\times$ 1$'$ field of the \textit{HST} image of the cluster, while the X-ray emission from this cluster mapped using \textit{Chandra} telescope appears as a single extended source covering about a few arcmin of the region centred on the core of ZwCl~2701. In addition to the extended emission, the hot gas distribution also provide evidences of depressions in the X-ray emission and are marked by arrows on the East and West directions of the X-ray peak of the cluster. To highlight these depressions in the X-ray image, we have used various techniques that are discussed below. 

%============================================
\begin{figure*}
\vbox
{
\includegraphics[scale=0.35]{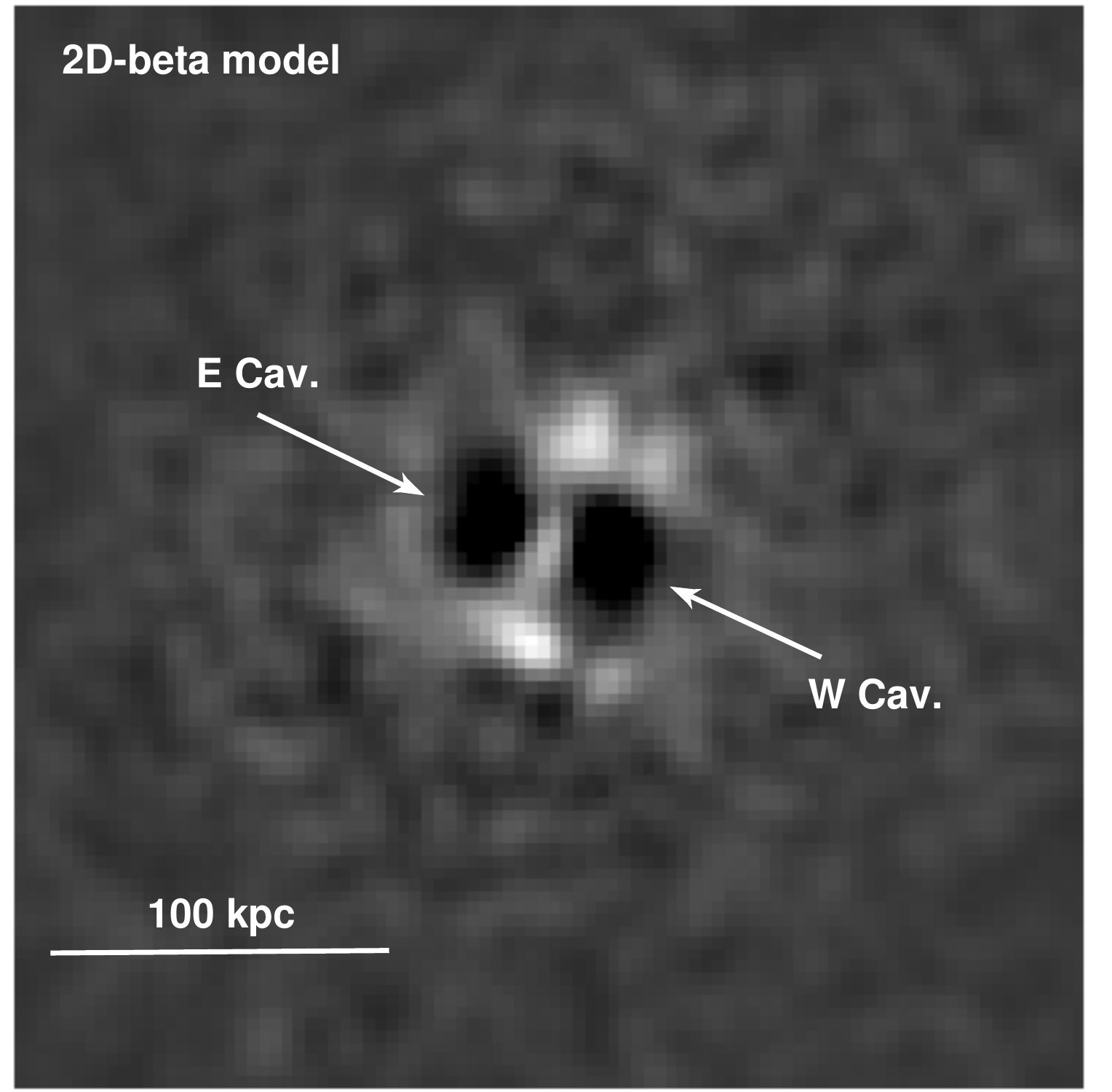}
\includegraphics[scale=0.35]{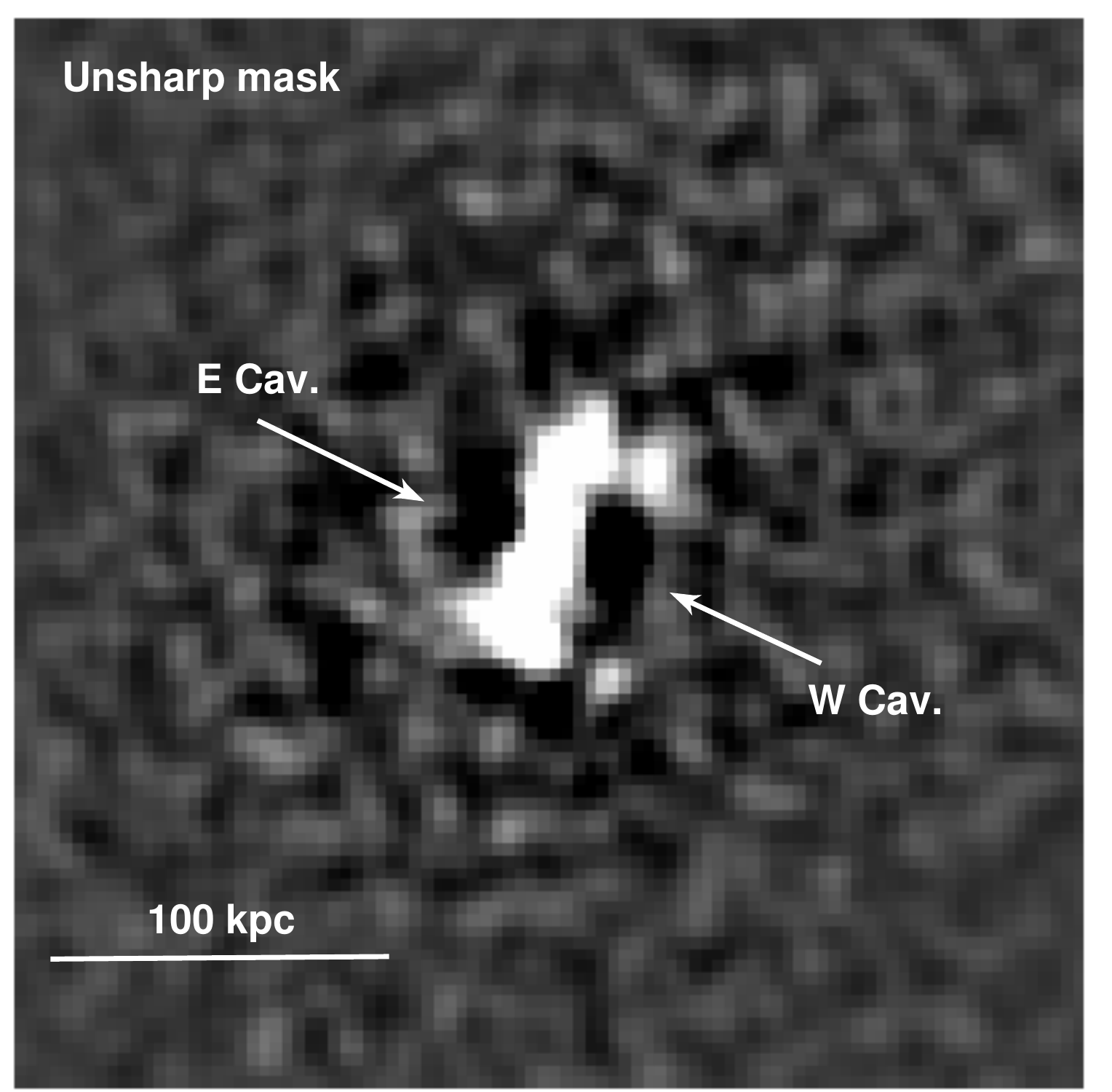}
\includegraphics[scale=0.35]{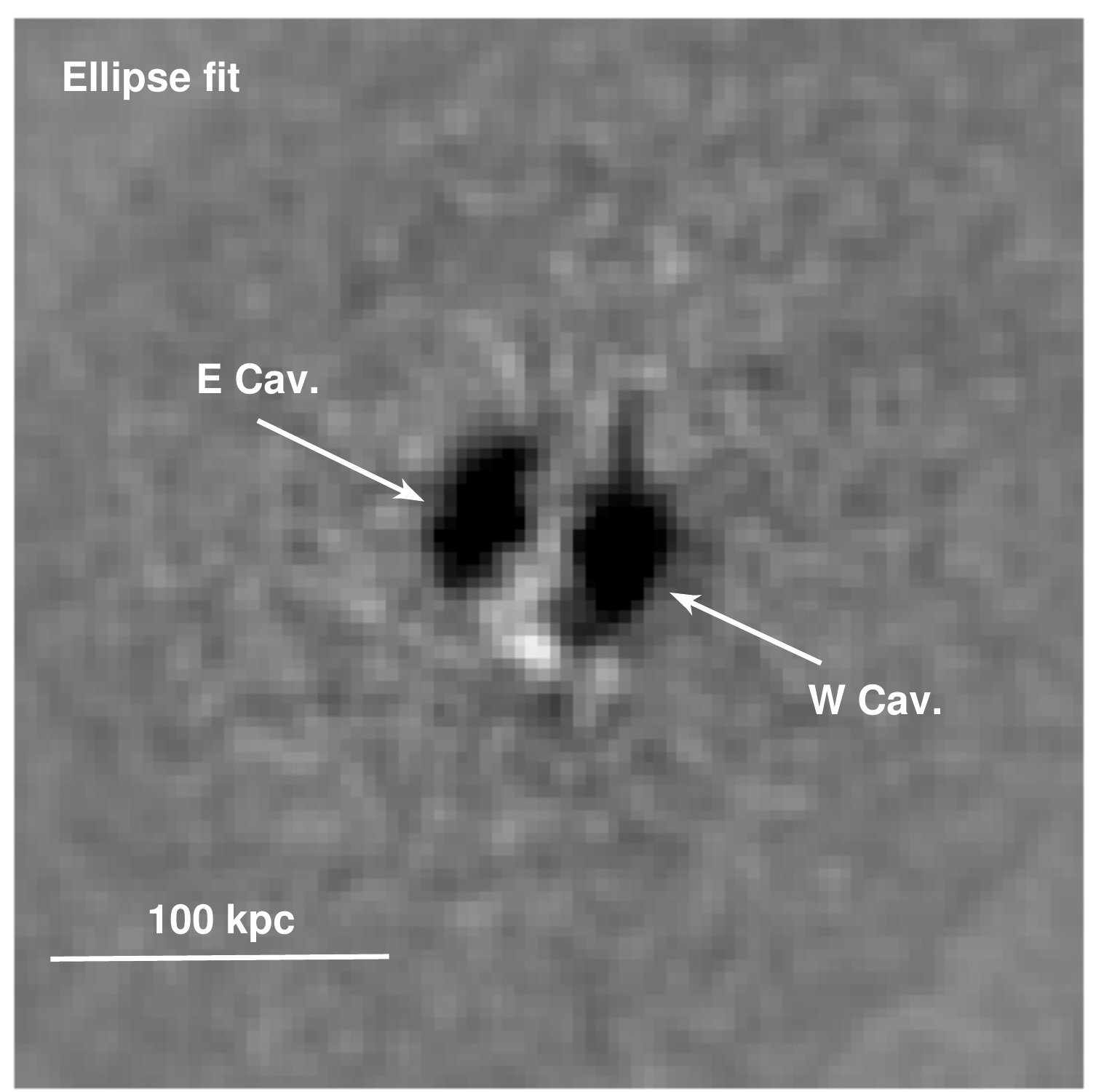}
}
\caption {\textit{Left panel:} \textit{Chandra} 0.3$-$3\,keV residual image of ZwCl~2701 generated 
after subtraction of the best fitted elliptical 2D-beta model and smoothed with a 2D Gaussian kernel 
of width 5 pixel ($\sim$ 2.5 arcsec). In the image, North is up and East is to left.
\textit{Middle panel:} \textit{Chandra} ACIS-S3 exposure corrected, background subtracted 0.3$-$3\,keV 
smoothed unsharp masked image. \textit{Right panel:} \textit{Chandra} 0.3$-$3\,keV residual image of 
ZwCl~2701 generated by using IRAF ellipse fit task. All the images are of 1.5\arcmin \tim 1.5\arcmin\, 
size and clearly reveal the presence of surface brightness depressions.}
\label{fig2}
\end{figure*}

%============================================

For the systematic investigation of X-ray cavities and other hidden features in the environment 
of ZwCl~2701, we have derived residual map after subtracting a smooth model from the original image 
(\citealt{2010ApJ...712..883D}), an unsharp mask image \citep{2009ApJ...705..624D} and residual 
image generated by using IRAF ellipse fit task (\citealt{2007A&A...461..103P,2012NewA...17..524V}). 
The 2D smooth model was derived by fitting 
2D-$\beta$ model to the isophotes in 0.3-3 keV \textit{Chandra} image of the cluster. The 
2D-$\beta$ model resulted into the best-fit parameters of slope $\alpha\, =\, 1.23 
\pm 0.02$, core radius $r_0\, =\, 13.55\arcsec \pm0.33$, ellipticity $\epsilon\, =\, 0.27 \pm 0.01$ 
and position angle $\theta\, =\, 1.04\pm0.02\, rad$. The smooth model was then subtracted from the 
original image and the resultant residual map is shown in {left panel of Figure~\ref{fig2}}. The 
unsharp mask image was obtained by subtracting a 6$\sigma$ Gaussian smoothed 0.3-3 keV image from 
that smoothed with 1$\sigma$ Gaussian image in same energy band and shown in the middle panel
of Figure~\ref{fig2}. For smoothing the images, we used the task \textit{aconvolve} within 
\textsc{CIAO}. Right panel of Figure~\ref{fig2} shows the residual image generated by 
using the IRAF ellipse task. The ellipse task fits the elliptical isophotes to galaxy image 
and gives the isophote intensity table file. Using this table file, a model image is being 
built and then subtracted from the original image. All these three images obtained from 
different techniques clearly reveal a pair of cavities or bubbles in the X-ray surface brightness 
distribution (darker shades) and appear to be surrounded by bright rim-like features. In addition 
to the pair of cavities, two bright knots (N-knot and S-knot) are also evident in these figures. 

%======================================================
\begin{figure*}
\hspace{-1cm}
\includegraphics[width=9.0cm,height=7.0cm]{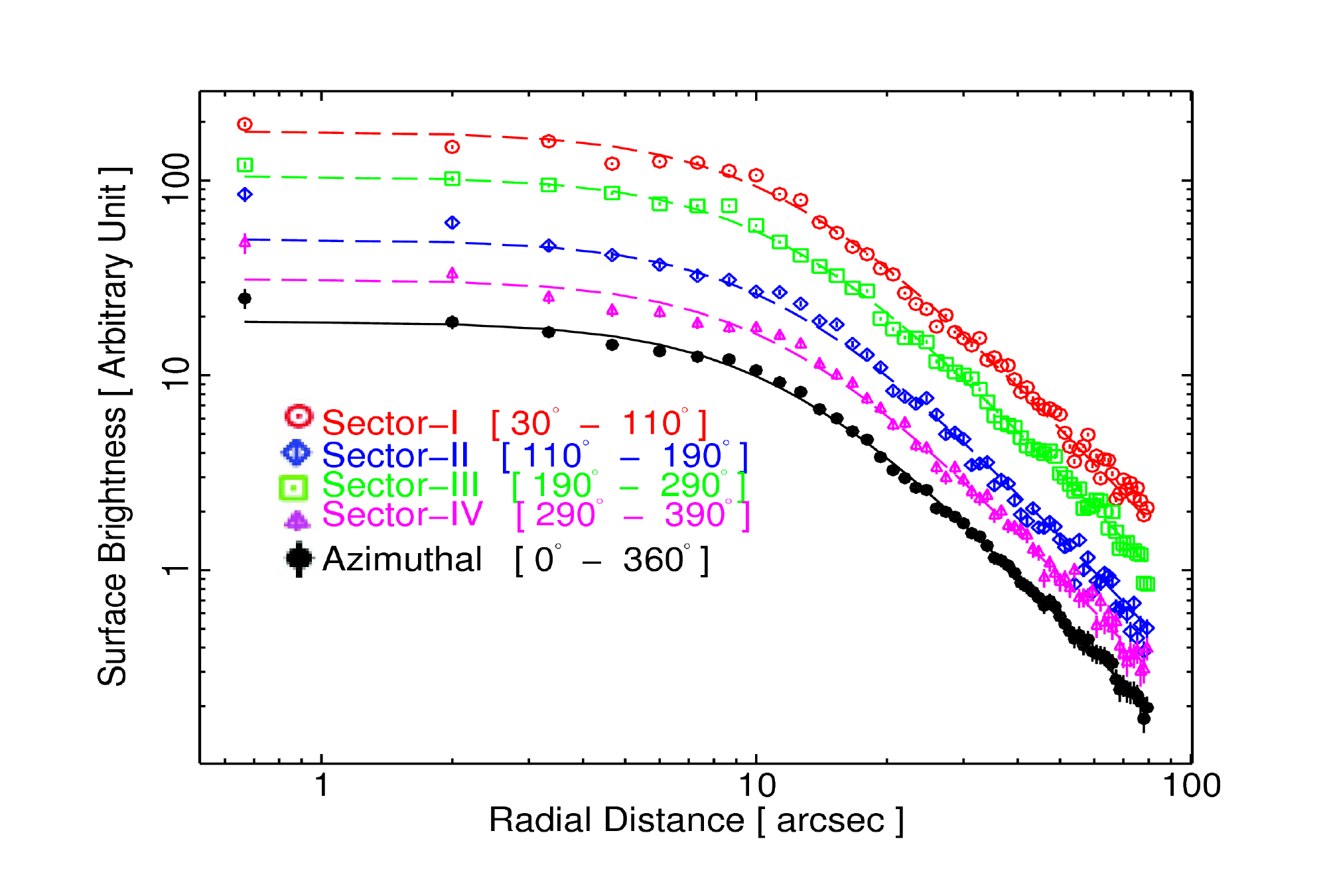}
\includegraphics[width=9.0cm,height=7.0cm]{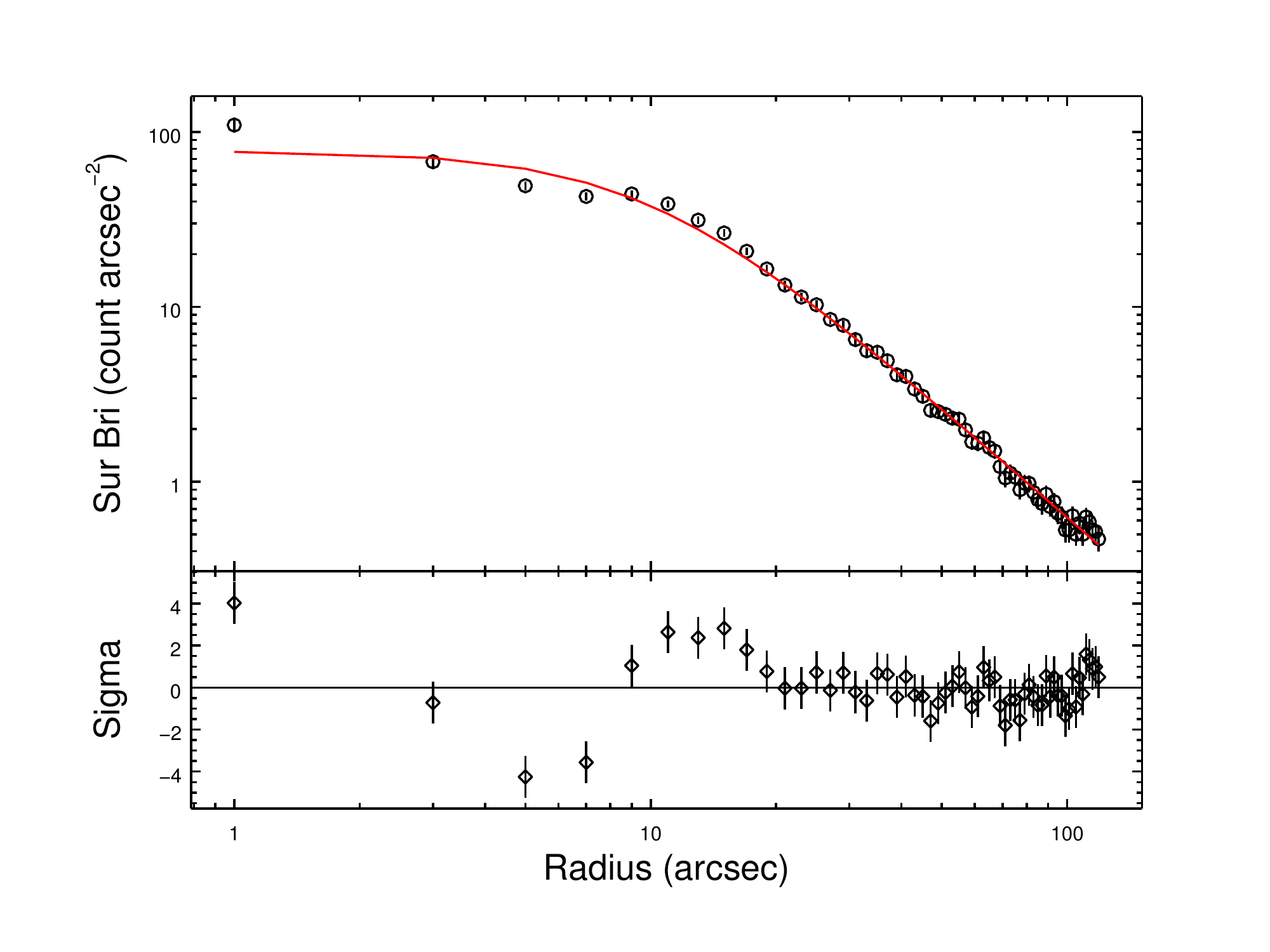}
\caption{{\it Left panel} : The projected azimuthally averaged, 0.3$-$3\,keV radial surface brightness profile fitted with 1D-beta model is shown as solid line in the figure (left panel). The Projected surface brightness profiles are derived from 4 different sectors, Sector I ($30^{\circ}$ - $110^{\circ}$), Sector II ($110^{\circ}$ - $190^{\circ}$), Sector III ($190^{\circ}$ - $290^{\circ}$) and Sector IV ($290^{\circ}$ - $30^{\circ}$). The profiles are offset by arbitrary values for clarity of presentation. For comparison, the best-fit azimuthally average 1D standard $\beta$ model is plotted as a dashed line. 
{\it Right panel}: The surface brightness profile of both the cavity regions added together is compared 
with non-cavity side 1D $\beta$-model in the top panel. The deviation of the surface brightness profile 
from the 1D $\beta$-model is shown in the bottom panel of the figure.}
\label{fig3} 
\end{figure*} 
%===========================================  

To investigate extent of the X-ray emission from this cluster, we have generated its surface brightness profile. This was done by extracting 0.3-3 keV X-ray photons from a series of concentric elliptical annuli, each of 2\arcsec\, bin-width, centred on the X-ray peak of the cluster upto 2\arcmin\, radius. The resulting background subtracted exposure corrected radial surface brightness profile is shown as solid line in Figure~\ref{fig3}. Assuming that the ICM and the galaxies in the cluster are in hydrostatic equilibrium, we fit this surface brightness profile with a  single $\beta$ model 
\begin{equation}
\resizebox{0.5\hsize}{!}{$\Sigma_{(r)} = \Sigma_{(0)} \left[1 + \left(\frac{r}{r_0}\right)^2 \right]^{-3\beta+0.5}$}
\end{equation}
where $\Sigma_{(0)}$, $\beta$ and $r_0$ represent the central brightness, the core radius and the slope, respectively. 
The fitted slop $\beta$ and core radius $r_0$ were found to be $\approx$ 0.51 and $\approx$ 9.75$\arcsec$ (33.95 kpc), respectively.

%=================================================================
\begin{figure*}
\centering
\includegraphics[scale=0.55]{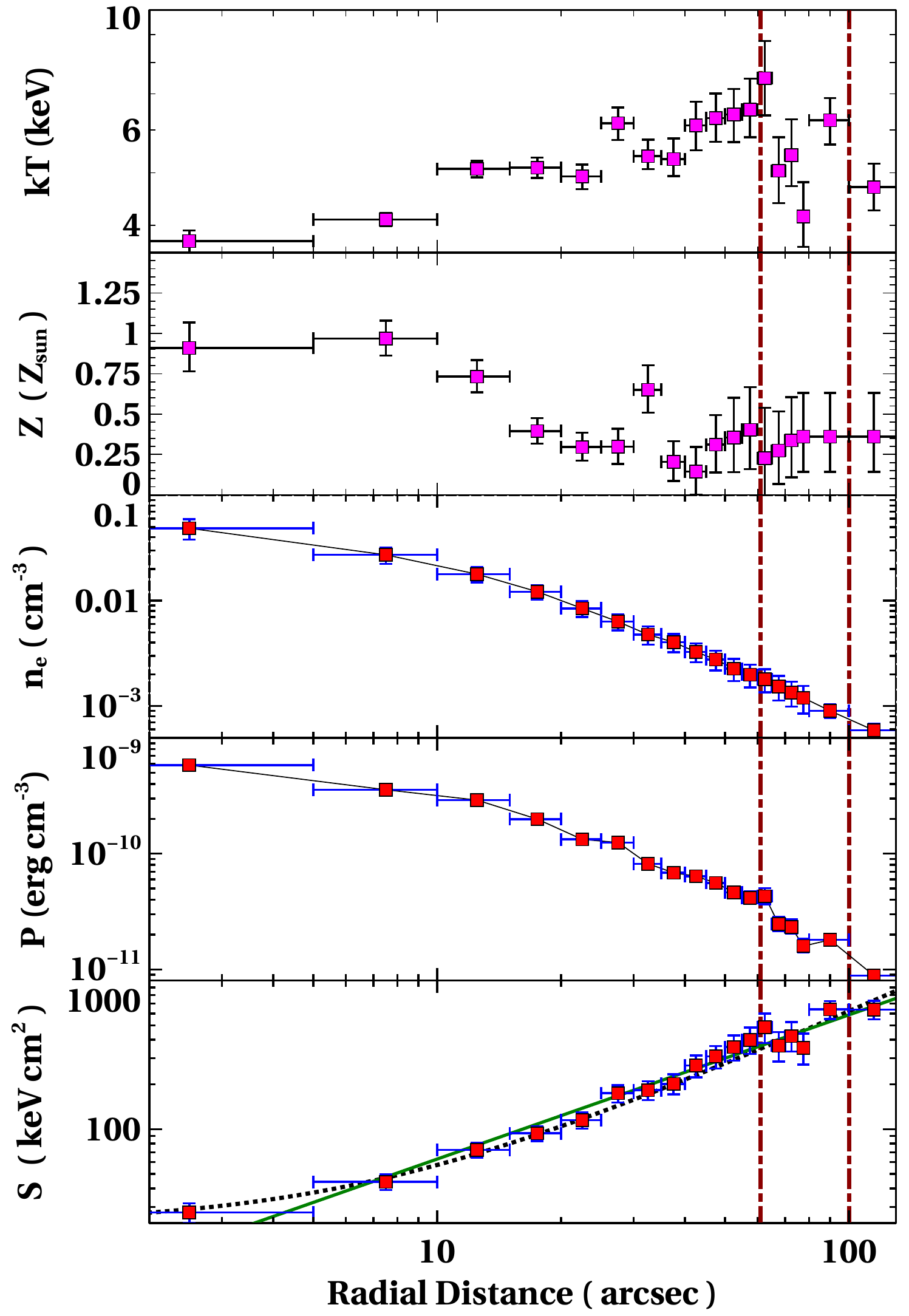} 
\caption{\label{spec} Azimuthally averaged projected temperature, metallicity, 
electron density, pressure and entropy profiles of ZwCl~2701 as a function of 
radial distance from the core of the cluster. Entropy profile with dashed line 
and solid line representing the best-fit of constant + power law model 
$S(r)=S_0 + S_{100} (r/100kpc)^{\alpha_1}$ and only power law component alone 
$S_{100} (r/100kpc)^{\alpha_2}$, respectively.}
\label{fig4} 
\end{figure*} 
%=================================================================

To elucidate these depressions, we derived similar radial surface brightness profiles for X-ray photons extracted from four different wedge shaped sectors covering  Sector I ($30^{\circ}$ - $110^{\circ}$), Sector II ($110^{\circ}$ - $190^{\circ}$), Sector III ($190^{\circ}$ - $290^{\circ}$) and Sector IV ($290^{\circ}$ - $30^{\circ}$). The resultant radial surface brightness profiles for above four sectors are also shown in Figure~\ref{fig3}. To better represent these profiles, arbitrary scales were applied on the surface brightness. To compare the distribution of estimated values of surface brightness profiles for each sector, the best fitted azimuthally averaged 1D-$\beta$ model with fixed slope and core radius was over plotted in the figure. The surface brightness plots showed that the gas distribution in ZwCl~2701 cluster is not symmetric in all directions but indicate the presence of marginal depressions in distribution along Sectors I \& III. 
Apart from this, the profiles shows an excess emission at 8\arcsec - 20\arcsec\, region and a highly 
complex nature above \s 20\arcsec\, radii. 

Despite the clear detection of cavities in the residual of 2D-$\beta$ model (Figure~\ref{fig2}), 
depressions seen in the surface brightness profiles obtained from all the four sectors (left panel of 
Figure~\ref{fig3}) are marginal. To investigate the presence of depressions, the surface brightness 
from both the cavity side regions were extracted and added together and then compared with that from
the non-cavity side 1D-beta model. The best fitted 1D-beta model parameters are found to be core radius r$_0$ = 10.03$^{+0.28}_{-0.27}$ arcsec and $\beta$ = 0.51$_{-0.004}^{+0.004}$. Combined surface brightness 
profile of cavity sides and the non-cavity side 1D-beta model are shown in the right top panel of 
Figure~\ref{fig3} whereas the deviation between them is shown in the right bottom panel. The presence 
of surface brightness depressions can be clearly seen in 3--7 arcsec radii from the core of the cluster 
and agrees with the residual images.

\subsection{X-ray Spectral Analysis}

To investigate thermodynamical properties of the ICM, we extracted 0.5-7 keV X-ray spectra from 24 
concentric elliptical annuli, each of width 5 \arcsec, centred on the X-ray peak of ZwCl~2701. 
Background spectrum was extracted from the blank sky background file. Response matrices and photon 
weighted effective area files were generated for each of the spectrum using the task \textit{specextract} 
of CIAO. After appropriate background subtraction, each spectrum was fitted independently assuming the 
collisionally-ionized diffuse gas (ATOMDB code) subjected to the Galactic photoelectric absorption model 
(\textit{WABS} $\times$ \textit{APEC}). The \textit{APEC} code has three free parameters i.e. plasma 
temperature (in keV), metal abundance (in solar units) and the normalization component (cm$^{-5}$). 
In spectral fitting, all the three parameters were allowed to vary while the value of the equivalent 
hydrogen column density in the source direction was kept fixed at the Galactic value (7.9 $\times\, 
10^{19}\,$ cm$^{-2}$).  Electron density $n_e$ of the gas was estimated from the \textit{APEC} 
normalization which is directly related to the density through the relation
\begin{equation}
N_{apec} = \frac{10^{-14}}{4\pi [D_A (1+z)]^2} \int\! n_e n_H \mathrm{d}x
\end{equation}
where, $D_A$ is the angular diameter distance to the source (cm), $n_e$ and $n_H$ are the electron 
and hydrogen densities (cm$^{-3}$). Here, we assume $n_e/n_H \sim 1.2$ for the solar abundance (for 
more details see \citealt{2012AdAst2012E...6G}). Once the values of temperature and density of the gas 
were found, the gas pressure {\it P} and entropy {\it S} were estimated by using relations \textit{P} = 
$nkT$ ($n$ being total density = 1.92 $n_e$) and $S=kT n_e^{-2/3}$. The error on density, pressure 
and entropy were calculated by using error propagation method described by \cite{2003drea.book.....B}.

%=======================================================================
\begin{figure*}
\includegraphics[scale=0.5]{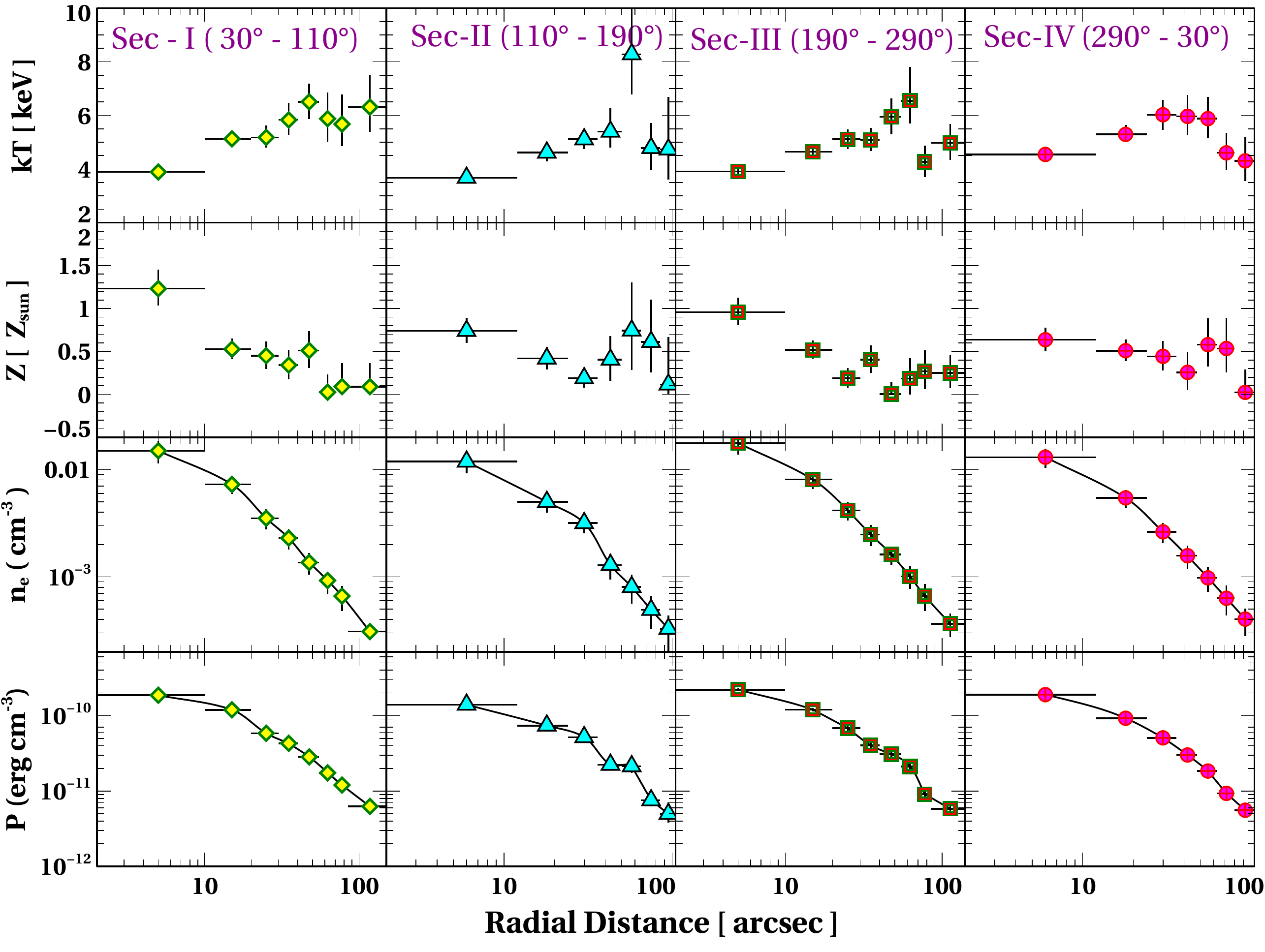}
\caption{Projected temperature, metallicity, electron density and pressure profiles derived for four
sectors. All spectral uncertainties are at 90$\%$.}
\label{fig5}
\end{figure*}
%========================================================================

The resulting radial profiles of temperature (\textit{kT}), metallicity (\Zsun),  electron density ($n_e$), 
pressure (\textit{P}) and entropy (\textit{S}) of the ICM from ZwCl~2701 are shown in top to bottom panels 
of Figure~\ref{fig4}, respectively. Estimated temperature, electron number density and gas pressure at the 
core of the cluster were found to be $\sim$3.8 keV, $\sim$0.05 cm$^{-3}$ and 0.35 keV cm$^{-3}$, respectively. 
As ZwCl~2701 is a cool core cluster, the value of the estimated temperature from the core gradually increases 
up to a maximum of $\sim$7.25 keV at $\sim$60\arcsec\, whereas the electron density and gas pressure profiles 
showed a monotonous decrease in the corresponding values. A small drop in the gas temperature and pressure 
profiles was evident in the region about 60\arcsec -- 90\arcsec\, from center of the cluster. However, the 
electron number density profile do not show any deviation. Similar nature of radial profiles of gas 
temperature, electron number density and pressure profiles were also reported for the system MS~0735.6+7421 
(\citealt{2005Natur.433...45M}).

We have also obtained projected temperature, abundance, electron density and pressure profiles of the X-ray 
photons extracted from the same four sectors as discussed above and the resultant profiles are plotted from 
left to right panels in Figure~\ref{fig5}. Except the temperature profiles along all the four sectors showed 
positive gradient as a function of radial distance from the core. A minor dip is evident in the temperature 
profiles along all the four sectors at $\sim$60\arcsec. However, the metal abundance profiles showed an 
overall negative gradient nature as expected in the cool core clusters. In general, all the profiles look 
similar to those expected for cool core clusters with the temperature dropping in the core almost by a 
factor of \s2 relative to the peak temperature at \s 100\arcsec. The metallicity profiles along sector 
II and IV are on-jet profiles whereas sector I and III are off-jet. The profiles along the on-jet side show 
rise in metallicity at larger radii (\s 60\arcsec- 80\arcsec). Our results on metallicity profiles of 
ZwCl~2701 along the on-jet side coroborates the results obtained for other clusters such as 2A~0335, 
Hydra~A and A~2052 (\citealt{2015MNRAS.452.4361K}). The elemental abundance present in BCGs is uplifted
due to the rising cavities (bubbles) from central twenty kiloparsec to an altitude of several hundred 
kiloparsec. This causes higher metallicity at larger radii.

To examine the global properties of the hot gas within ZwCl~2701, source spectrum in 0.3-7\,keV range
was extracted from a circular region of 2\arcmin\, centred on the X-ray peak. The spectrum was then 
fitted with a simple collisionally ionized plasma model with the photo-electric absorption fixed at 
the Galactic value. The best-fit parameters of the spectrum are listed in Table~\ref{tab1}.

%==========================================================
\begin{table*}
\caption{Thermodynamical parameters of the ICM from the regions of interest in ZwCl~2701}
\vbox{%
\centering
\begin{tabular}{@{}cccccccr@{}}
\hline
Reg. & kT (keV)/$\gamma$ & Z (Z$_{\odot}$) &  L$_{0.3-10keV}$ (10$^{42}$ erg s$^{-1}$) & counts & $C-stat$ & $\chi^2/dof$ \\
\hline
2'($\sim$ 420 kpc)     & 5.14 $\pm$ 0.12 & 0.36 $\pm$ 0.04 & 678 $\pm$ 04    & 68661    & --         & 359.00$/$354 \\
Clumpy                 & 4.11 $\pm$ 0.16 & 0.54 $\pm$ 0.07 & 165  $\pm$ 2.00 & 16183    & --         & 209.09$/$216 \\
E Cav.                 & 3.47 $\pm$ 0.50 & 0.70 $\pm$ 0.40 & 6.99 $\pm$ 0.04 & 707      & 57.53$/$56 & -- \\ 
W Cav.                 & 5.36 $\pm$ 0.90 & 0.94 $\pm$ 0.50 & 10.1 $\pm$ 0.03 & 875      & 63.08$/$67 & -- \\
S Knot                 & 4.01 $\pm$ 0.52 & 0.74 $\pm$ 0.36 & 10.1 $\pm$ 0.02 & 965      & 75.31$/$76 & -- \\
N Knot                 & 4.10 $\pm$ 0.58 & 0.66 $\pm$ 0.35 & 9.95 $\pm$ 0.22 & 951      & 60.48$/$71 & -- \\
Nucli                  & 3.52 $\pm$ 0.61 & 0.32 $\pm$ 0.30 & 3.66 $\pm$ 0.30 & 354      & 26.70$/$29 & -- \\
                       & 1.97 $\pm$ 0.11 &     --          &    --           & 354      & 29.92$/$30 & -- \\
\hline
\end{tabular}}
\label{tab1}
\end{table*}
%======================================================

\begin{figure*}
\centering
\includegraphics[width=7cm, height=7cm]{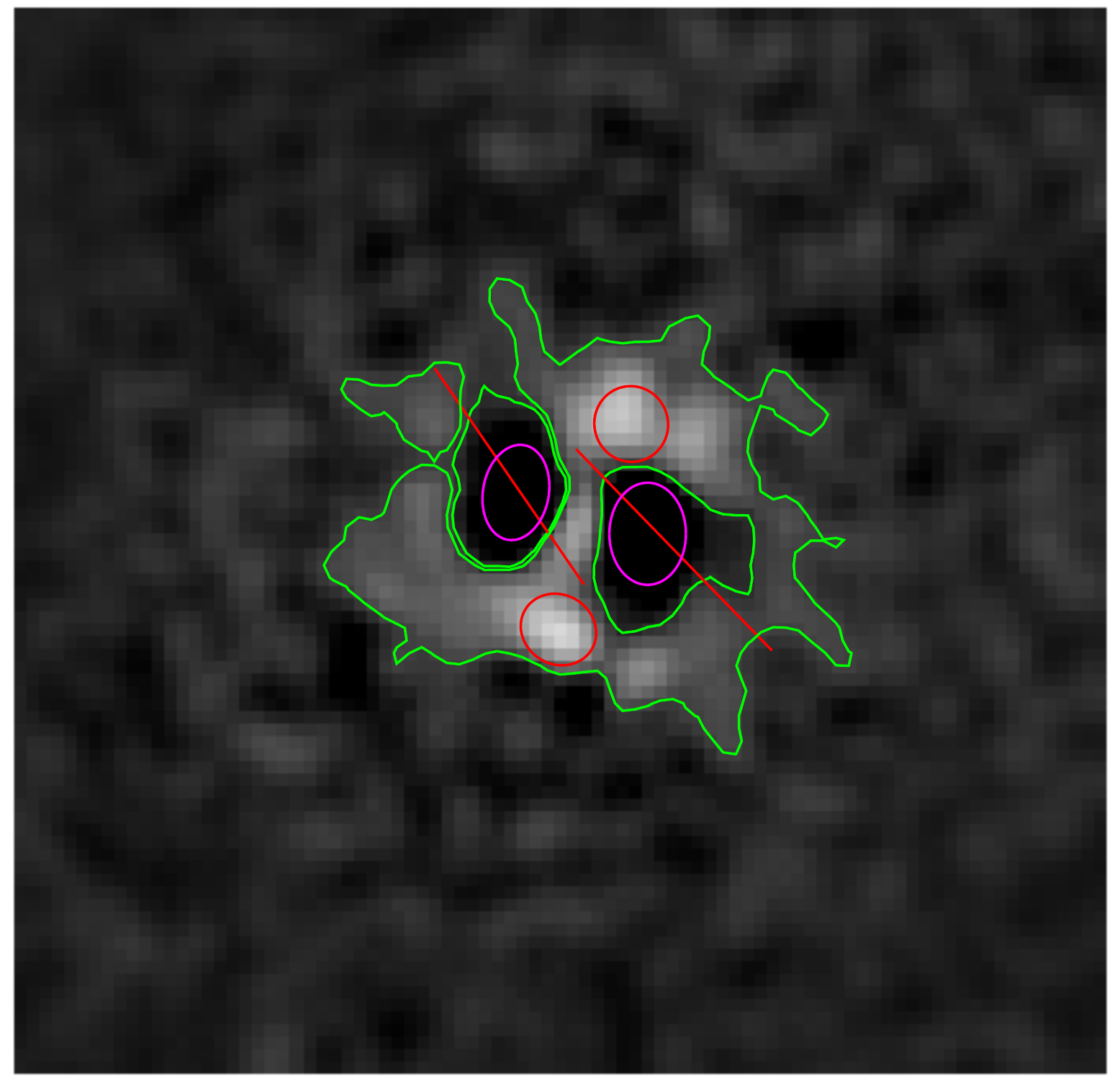}
\caption{ Residual image showing regions with interesting features such as cavities (magenta
ellipsoids), knots (red circles) and clumpy regions (green contours excluding cavities).  
The ellipsoids drawn in the figure are scaled with the cavity sizes as quoted in Table~2.} 
\label{fig6} 
\end{figure*}

We have also performed spectroscopy by extracting source spectrum from the regions of interest separately. Several interesting features such as cavities, clumpy regions, knots and nucleus were seen in the residual image of ZwCl~2701 (Figure~\ref{fig6}). The X-ray deficient cavity regions are shown by dark shades while the knots (excess emission) appeared as the bright regions in the figure and are highlighted with magenta ellipses and red circles, respectively. As the shape of each of the feature was irregular, it was difficult to extract source spectrum for each of the features accurately. To study these feature independently, we have extracted source spectra from these regions separately. The spectrum for the clumpy region was obtained by extracting X-ray photons from the regions shown by green contours and excluding cavity regions (Figure~\ref{fig6}). Spectra extracted from these regions of interest were then fitted using APEC model within XSPEC. The results from the best-fit spectra of clumpy, cavity regions and the knots are presented in Table~\ref{tab1}.

%=================================================
%=================================================

To investigate nature of the nuclear source associated with this cluster, we extracted a 0.3 - 8 keV spectrum from within the central 1.5\arcsec\, region centred on RA=9:52:49.198, DEC=+51:53:05.19. The central source appears to be embedded within the diffuse gas, which may contaminate the hard spectral component. We fitted this spectrum with a simple power law and a single temperature \textsc{APEC} model independently, with the Galactic absorption ($n_H$) fixed. The best fit power-law component yielded the photon index $\gamma$ \s $1.97^{+0.11}_{-0.11}$ whereas the single temperature model yielded gas temperature kT = $3.22^{+0.52}_{-0.71}\, keV$ and metal abundance $Z = 0.32^{+0.23}_{-0.41}\, Z_{\odot}$ (Table~\ref{tab1}). X-ray luminosity of the nuclear source was found to be $L_{2-10keV}$ $\sim$ 2.27 \tim $10^{42}$\, \lum. 

\subsection{Temperature and abundance maps}

%=======================================================================
\begin{figure*}
\vbox{
\includegraphics[trim=00 1.2cm 00 00, width=70mm,height=70mm]{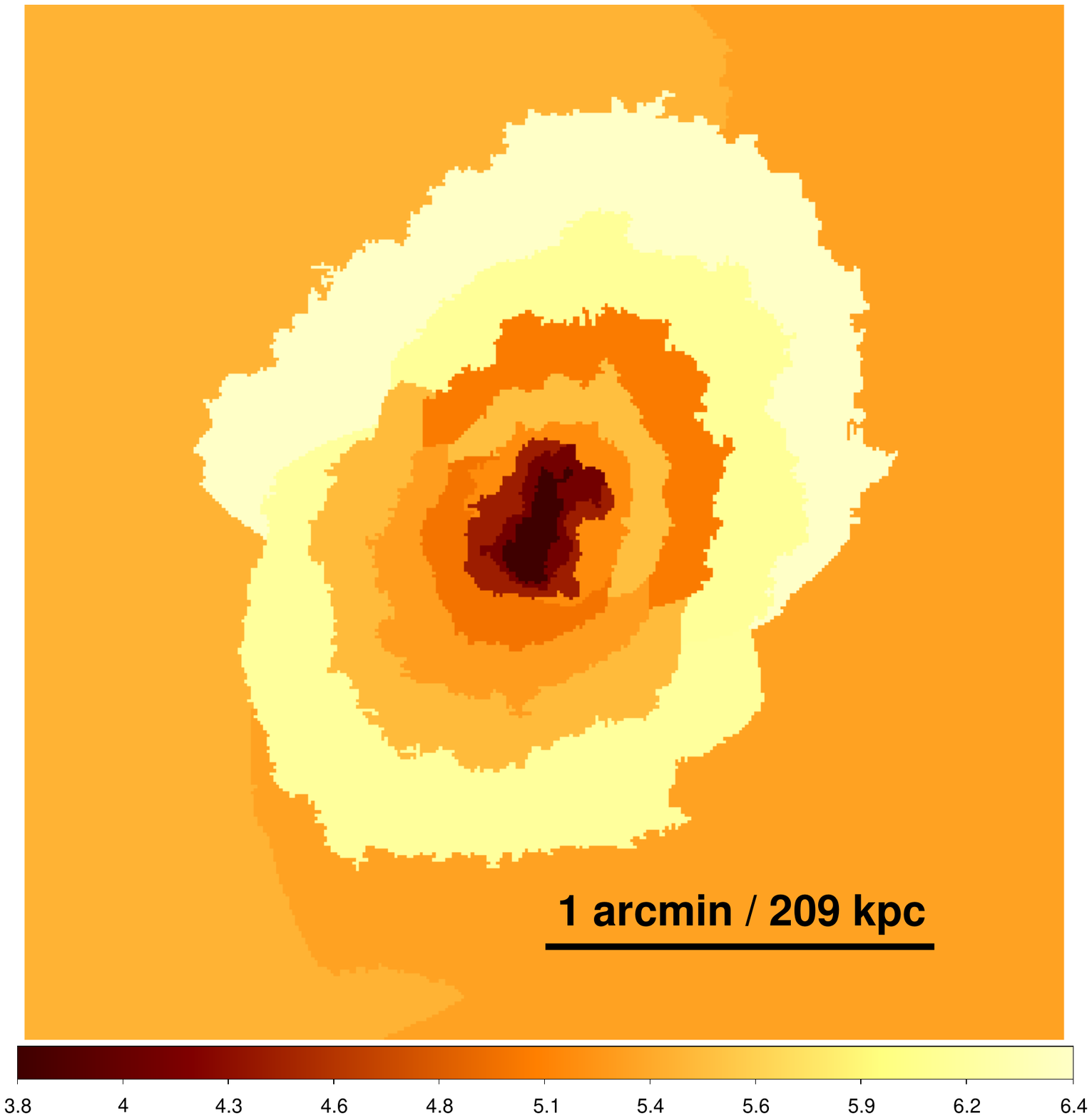}
\includegraphics[trim=00 1.2cm 00 00, width=70mm,height=70mm]{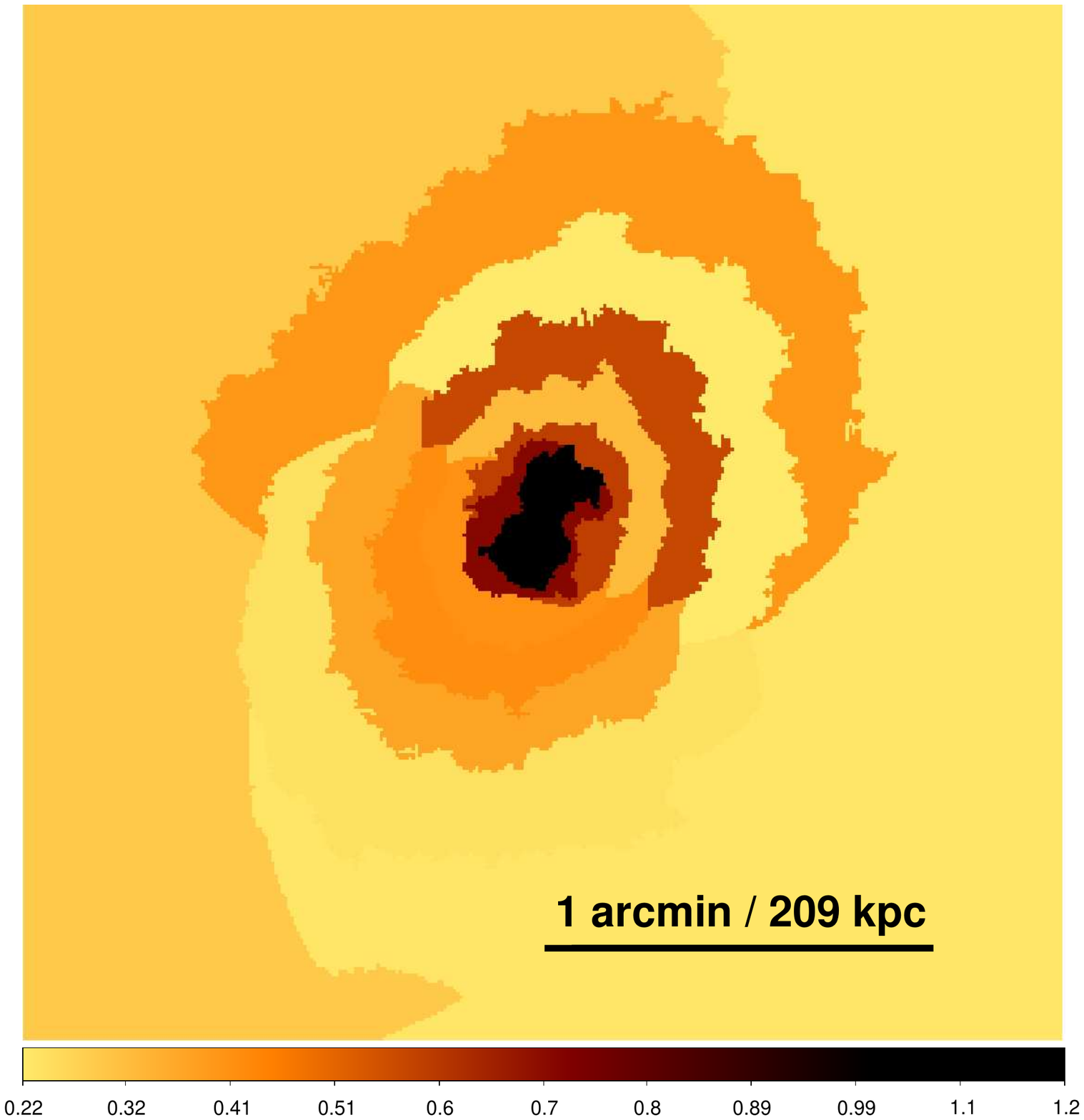}
}
\caption{Projected temperature map (left panel) in units of keV and metallicity map (right panel) in 
units of $Z_{\odot}$ generated by using contour binning technique with large spectral bins having 
signal-to-noise ratio of 60 ($\sim$ 3600 counts). Bright and darker shades in the temperature map 
indicate hotter and cooler regions, respectively, whereas in abundance map, darker shade represent 
higher metallicity.}
\label{fig7}
\end{figure*}
%========================================================================

The superb spatial resolution capability of \textit{Chandra} was used to investigate the temperature and metallicity distribution across the cluster ZwCl~2701. Therefore, we used the central 160\arcsec \tim 160\arcsec\,(558 \tim 558 kpc) image to construct the temperature and abundance maps of the ICM within ZwCl~2701. Contour binning algorithm of \citet{2006MNRAS.371..829S} was used for generating these maps. This algorithm generates contours on an adaptively smoothed map and bins the X-ray data according to the surface brightness distribution in such a way that each of the selected region has at least \s 3600 counts (S/N \s 60). The spectra and appropriate response files were generated for each of the contour. As discussed earlier, the blank sky files were used for the background estimation. Each of the contour spectrum was grouped in such a way that it contained at least 25 counts per bin. The extracted spectra were then fitted with a simple APEC model with the fixed Galactic absorption and varying temperature, abundance and normalization parameters. The resulting temperature and metallicity maps of the contour binning are shown in Figure~\ref{fig7}. 

Temperature map ({\it left panel} of Figure~\ref{fig7}) showed that the gas in the core of ZwCl~2701 cluster is coolest (\s 3.8 keV) whereas maximum temperature of \s 6.4 keV was detected along the North direction at about 60\arcsec. From the metallicity map (right panel of Figure~\ref{fig7}), it is apparent that the highest metallicity (\s 1.2 \Zsun) is seen in the central region of the cluster which then decreases up to \s 0.22 \Zsun in the outward direction. Thus, the gradual rise in temperature and fall in metallicity (Figure~\ref{fig5}) are confirmed in the temperature and metallicity maps. The lower temperature and higher metallicity at the core relative to the outer region -- typical characteristics 
of the cool core clusters, have been noticed in ZwCl~2701.

\subsection{Radio and X-ray imaging}

%============================================
\begin{figure*}
%\centering
\hbox{
\includegraphics[clip, trim=0.75cm 4.5cm 0.5cm 2.5cm, scale=0.45]{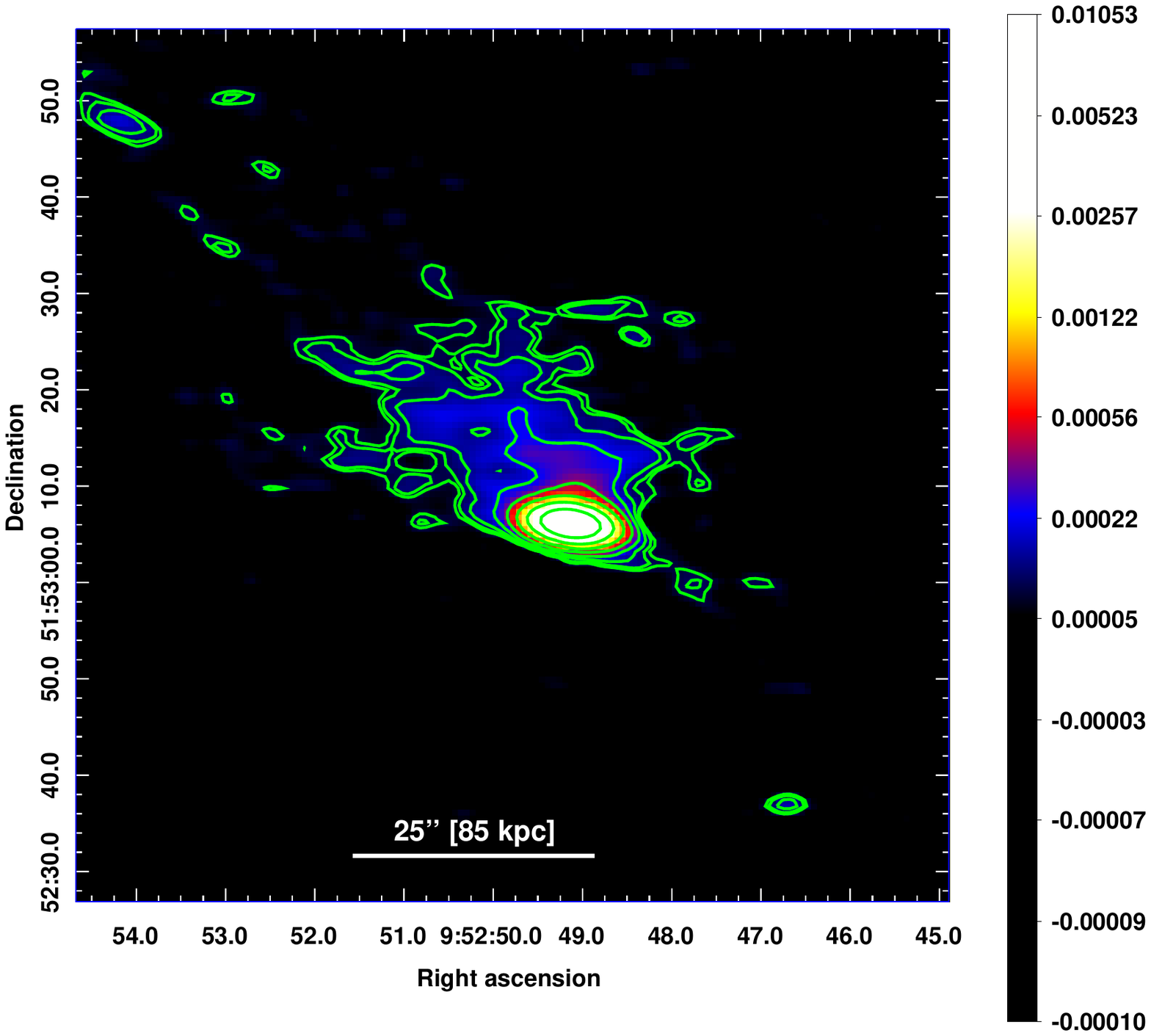}
\includegraphics[clip, trim=0.5cm 4.5cm 0.5cm 1cm, scale=0.45]{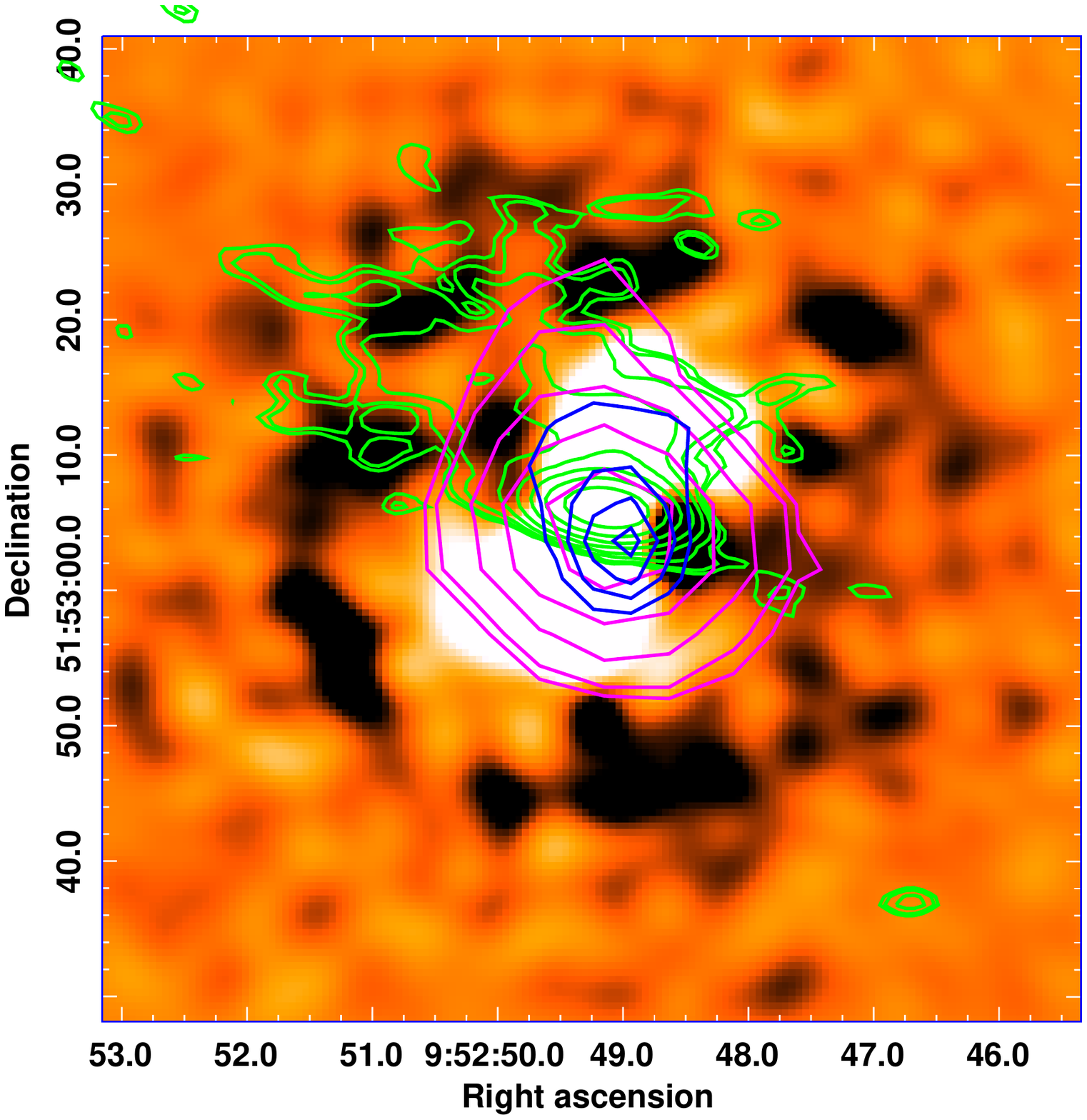}
}
\caption{ {\it Left panel:} \textit{GMRT} 1.4\,GHz radio emission map of ZwCl~2701. The 
image is smoothed with a Gaussian radius of 1 pixel. The restoring beam size is 5.52\arcsec 
$\times$ 2.4\arcsec, PA = 84$^{\circ}$ and the rms noise is 30 $\mu$Jy $beam^{-1}$. The 
3$\sigma$ logarithmic contour levels are $10^{-4}\, \times\, $(0.90, 1.05, 2.23, 4.20, 
8.87, 19.96, 46.25, 108.60) Jy. {\it Right panel:} \textit{Chandra} unsharp masked image 
with the overlaid 1.4\,GHz GMRT radio contours (green). For comparison, we also overlay 
the \textit{VLA}, 4.89\,GHz radio contours (magenta) and \textit{VLA} 8.49\,GHz radio 
contours (blue). The beam size for VLA 4.89\,GHz is 13\arcsec $\times$ 13\arcsec, the 
rms noise 1$\sigma$ = 130 $\mu$Jy $beam^{-1}$, and the 3$\sigma$ square root contours 
at 0.39, 0.53, 0.93, 1.62, 2.58, 3.81 mJy. The beam size for VLA 8.49\,GHz is 7.89\arcsec  
$\times$ 7.89\arcsec with the rms noise 1$\sigma$ = 136 $\mu$Jy $beam^{-1}$, and the 
3$\sigma$ linear contour levels at 0.41, 0.76, 1.10, 1.45, 1.80 mJy.}
\label{fig8}
\end{figure*}
%============================================

Multi-frequency radio observations are proxy to understand the AGN feedback activities in the 
cores of galaxy clusters \citep{2008ApJ...686..859B}. Radio jets originating from the AGN are 
believed to inflate bubbles in the ICM. Radio observations of the cool core clusters have 
demonstrated that the radio emission from the lobes fills the X-ray cavities and hence confirm 
their ubiquitous nature \citep{2004ApJ...607..800B,2006ApJ...652..216R,2010ApJ...720.1066C}. 
Therefore, it is important to find a link between them by investigating their association in 
ZwCl~2701. With these objectives, we have observed ZwCl~2701 at 1.4\,GHz with the \textit{GMRT}, 
Pune. 1.4\,GHz radio image with 2\arcsec.82$\times$ 1\arcsec.27 resolution (shown in 
Figure~\ref{fig8} \textit{left panel}) delineates the extended morphology of the radio 
emission that is elongated along the North-East and South-West direction. This confirms 
the center of the radio source in ZwCl~2701 coinciding with the X-ray peak and also with 
the brightest central galaxy. To compare the X-ray and radio emission features, we used 
the X-ray unsharp image of ZwCl~2701 cluster and overlay 1.4\,GHz radio emission contours. 
The resultant image is shown in the right panel of Figure~\ref{fig8}. The extended nature 
of the radio emission appears to trace both the X-ray cavities. The maximum distance of 
the radio jet from the radio peak is evident along the North-East direction and is \s 
59\arcsec\,(206 kpc). Further, elongation of the radio emission along the North-East 
direction exhibits its resemblance with the higher temperature gas seen in the 
temperature map (Figure~\ref{fig7}). 

Total flux density and flux density of core of ZwCl~2701 at 1.4\,GHz frequency were estimated 
to be \s 119 mJy with rms = 0.03 mJy and \s 11.30 mJy with rms of 0.04 mJy, respectively. 
\cite{2008ApJ...686..859B} using VLA multi-frequency radio data (except at 1.4\,GHz) estimated 
the integrated radio luminosity of ZwCl~2701 to be $\sim\, 4.0 \times 10^{41}\,$ \lum\,, where 
they assume a power-law nature $S_\nu\, \propto\, \nu^{- \alpha}$, $\alpha$ being the spectral 
index $0.7\pm0.1$. Our estimate of total radio luminosity of this source at 1.4\,GHz using 
\textit{GMRT} observations is found to be $\sim$ $2.20\, \times 10^{41}\,$ \lum.

\section{Discussion}
\subsection{AGN Feedback}
To estimate the total amount of energy (enthalpy) injected by the radio jets in to the ICM, 
we measured energy content of each of the cavity and the work done by them on their surrounding 
following (\citealt{2004ApJ...607..800B, 2006ApJ...652..216R})
\begin{equation}
E_{cav} = \frac{1}{\gamma_1 - 1}pV + pV = \frac{\gamma_1}{\gamma_1 - 1}pV
\end{equation}
where {\it p} and {\it V} are the surrounding gas pressure and volume of each of the cavity, 
respectively. X-ray cavities in the 2D-beta model subtracted residual image appeared as 
well-defined prolate ellipsoidal systems. Therefore, volume of the cavities were estimated by 
measuring their locations and sizes in the residual map as {\it V}=$4 \pi R_w^2 R_l/3$, where 
$R_l$ is the semi-major axis and $R_w$ is semi-minor axis. Though we adopted ellipse fitting 
method to show the presence of X-ray cavities in this cluster, it was important to determine the cavity 
size appropriately. We followed visual inspection method to determine the position and size 
of the cavity as used by \cite{2004ApJ...607..800B} for several clusters. Apart from the
visual inspection, the sectorial radial profiles of ZwCl~2701 (right panels of Figure~\ref{fig3}) 
provide a hint to determined the semi-minor axis of the cavity. The semi-major axis of the
cavity, however, can be estimated by extracting the counts from annular region around the 
depression. The uncertainty involved in determining the size of cavities by above method
contributed towards the estimation of cavity enthalpy. Here, we assume that both the cavities are 
filled with relativistic plasma of $\gamma_1$ = 4/3 so that the total enthalpy release to 
be $E_{cav} = 4{\it pV}$. The gas pressure of the cavities in the surrounding region
($p=nkT$ where $n$=1.92$n_e$) was obtained from the temperature and density profiles of the 
plasma. Once we have pressure and volume of the cavity, its enthalpy content can be 
derived, though it is difficult to estimate the error associated with it. This is because 
measuring size (or volume) of the cavity itself has certain uncertainty. The error 
measurement on the enthalpy reported in Table~\ref{tab2} is only due to the pressure.
The estimates of projected centroids, major and minor axes, cavity volume, density, 
temperature and enthalpy are reported in Table~\ref{tab2}. 

The total power injected by the radio jets into the surrounding medium was estimated 
by dividing the total enthalpy content of the cavities ($E_{cav}$) by their ages. The cavity 
power was derived by assuming all the three methods discussed by \cite{2004ApJ...607..800B} 
which are, $i.$ the time taken by the cavities to rise to the present position from the core 
at the sound crossing time ($t_{sonic}$); $ii.$ the time required for the cavities to rise 
buoyantly at their terminal velocities ($t_{buoy}$); and $iii.$ the time required to refill 
the displaced volume of gas as bubble rise in upward direction ($t_{refill}$). The cavity 
ages estimated by all the three methods are listed in Table~\ref{tab2} and lies in the range 
between \s 2.03 - 9.42 $\times$ 10$^7$ yr (see \citealt{2013Ap&SS.345..183P} for more detail). 
The total jet power required to create a cavity pair is simply P$_{jet}$ = E$_{cav}$/t$_{age}$, 
where E$_{cav}$ is the total cavity pair enthalpy and t$_{age}$ is the age of the cavity. For 
convenience, we assumed buoyancy age estimation method for the estimation of age of the cavities.
While age estimated using sound crossing time and refill time are the upper and lower limit to 
the age of the cavity, respectively (\citealt{2004ApJ...607..800B}). The total jet power is 
calculated to be P$_{jet}$ = 2.27$^{+0.28}_{-0.29}$ \tim $10^{45}$ erg s$^{-1}$ with upper and 
lower limits of 4.57$^{+0.56}_{-0.58}$ \tim $10^{45}$ erg s$^{-1}$ and $1.10^{+0.13}_{-0.14}$ 
\tim $10^{45}$ erg s$^{-1}$, respectively.

The presence of X-ray cavities in Zwcl~2701 cluster has been earlier reported by 
\cite{2006ApJ...652..216R} in a broad study of a sample of 31 groups and cluster galaxies. 
In that study, {\it Chandra} archival data of significantly less exposure (\s 27 ks) were
used, while the present study reports analysis of {\it Chandra} \s 123 ks exposure data. As 
mentioned in previous paragraph, the accuracy in the measurement of cavity size depends 
on the signal-to-noise ratio (exposure time) of the data. Considering this point, our 
estimation of cavity size that leads to the calculation of enthalpy, jet power are more 
reliable. Apart from these estimation from {\it Chandra} X-ray data, we also used {\it GMRT} 
1.4 GHz radio observations of the cluster to investigate the cavity position. Therefore, our 
estimation of cavity size and thermodynamical parameters are relatively more accurate 
compared to those reported earlier. 
%===================================== 
\begin{table}
\centering
\caption{\label{tab2} Energetic parameters of cavities}
\begin{tabular}{@{}cccr@{}}
\hline
Parameters &  E cavity & W cavity \\
\hline
$R_l$ (kpc) & 12.25 & 14.0 \\
$R_w$ (kpc) & 8.75  & 10.5 \\
$R$ (kpc) & 18.9  & 19.25 \\
Vol ($10^{68}\,$ cm$^{3}$) & 2.67 & 3.73 \\ 
kT (keV) & 4.23$^{+0.02}_{-0.02}$ & 4.79$^{+0.05}_{-0.05}$\\ 
$n_e$ ($10^{-2} \,$ cm$^{-3}$) & 9.62$^{+0.48}_{-0.45}$ & 9.01$^{+0.34}_{-0.30}$ \\
%$S$ (keV\, cm$^2$) & 19.77 & 42.27 \\
$p$ ($10^{-9}$\, erg cm$^{-3}$) & 1.25$^{+0.12}_{-0.12}$ & 1.34$^{+0.18}_{-0.18}$ \\
E$_{cav}=4pV$ ($10^{60}\,$ erg) & 1.14$^{+0.12}_{-0.11}$ & 1.99$^{+0.28}_{-0.27}$ \\
$C_{sound}$ (km\,s$^{-1}$) & 945 & 1175 \\
$v_{cavity}$ (km\, s$^{-1}$) & 513 & 554 \\
$t_{sonic}$ ($10^7\,$ yr) & 2.46 & 2.03 \\
$t_{buoy}$ ($10^7\,$ yr) & 4.52 & 4.30 \\
$t_{refill}$ ($10^7\,$ yr) & 8.59 & 9.42 \\
%$P_{sonic}$ ($10^{44}\,$ \lum)  & 6.74 & 17.52 \\
%$P_{buoy}$ ($10^{44}\,$ \lum)  & 3.66 & 8.26 \\
%$P_{refill}$ ($10^{44}\,$ \lum)  & 1.93 & 3.78 \\

\hline
\end{tabular}
%\end{center}
\footnotesize
\end{table}
%================================================== 

\subsection{Central engine in ZwCl~2701}
As the central engine is responsible for the formation of X-ray cavities and other 
substructures evident in the cluster ZwCl~2701, it is important to probe properties 
of the central engine. In this regard, estimation of the black hole mass in the central 
dominant galaxy of the cluster is one of the most important aspect (\citealt{2013MNRAS.431.1638H}) 
and can be estimated by using the well established relation between the black hole mass and the 
central stellar dispersion velocity $M_{BH}-\sigma$ (\citealt{2002ApJ...574..740T}). For ZwCl~2701, 
the stellar dispersion velocity was estimated by using the  \textit{SDSS} spectrum and was found 
to be \s 283 km s$^{-1}$. The $M_{BH}-\sigma$ relation lead to the central black hole mass of 
$\sim 5.44 \times 10^8 M_{\odot}$ and is consistent with $M_{BH,L_K}=6 \times 10^8 M_{\odot}$ 
reported by \citet{2006ApJ...652..216R} using the K-band luminosity. It is understood that the 
accreted mass on to the SMBHs is responsible for the AGN outburst, creating X-ray deficiency 
regions or cavities. We estimated mass accretion rate $\dot{M}_{acc}$ on to the SMBH in central 
dominant galaxy of ZwCl~2701 by using the relation suggested by \cite{2006ApJ...652..216R} and 
found to be $ 0.11 $ \Msun yr$^{-1}$. 

In the case of a fully ionized gas, the Eddington accretion rate $\dot{M}_{Edd}$ is a function 
of the black hole mass and radiative efficiency $\epsilon$ and is given by $\dot{M}_{Edd} 
= L_{Edd}/\epsilon\,c^2$, where, $L_{Edd}$ is the Eddington luminosity. Here we assume 
$\epsilon$ = 0.1 for the present calculation though its upper limit for the radiative 
efficiency is $\epsilon\, \leqslant$ 0.06 for a non-rotating black hole \citep{2002apa..book.....F} 
and that for an extreme Kerr black hole, it is $\epsilon\, \leqslant$ 0.4 (\citealt{2006ApJ...652..216R}). 
For a black hole mass of 5.44 $\times 10^8 M_{\odot}$ as in the case of ZwCl~2701, the Eddington 
luminosity is $L_{Edd} = 1.26 \times 10^{38} (M_{BH}/M_{\odot})\,$ erg s$^{-1}$ = $ 6.85 \times 
10^{46}$ \lum. This lead to $\dot{M}_{Edd} \sim 12\,$ \Msun yr$^{-1}$. The ratio of 
$\dot{M}_{acc}/\dot{M}_{Edd} = 9.16 \times\, 10^{-3}$ is below the threshold limit of 
0.01$\dot{M}_{Edd}$ required for the accretion flow to be radiatively inefficient 
(\citealt{1994ApJ...428L..13N}). This suggests that the launching of powerful jets 
from the SMBH in ZwCl~2701 is efficient when the accretion flow in the vicinity of 
the black hole has large scale height (\citealt{2012MNRAS.424.2971O}).

We also calculated the Bondi accretion rate of the SMBH defined as a function of the 
ambient gas density and the black hole mass (\citealt{1952MNRAS.112..195B}). We 
estimated the gas density by analysing the X-ray spectrum extracted from the central 
1.5\arcsec\, region and was found to be $n_e = 0.10\,$ cm$^{-3}$. The Bondi accretion 
process occurs within Bondi radius of $R_B = 2G M_{BH}/c_s^2$ under the influence of 
black hole, where $c_s$ is the sound speed in the surrounding medium and $M_{BH}$ is 
the black hole mass. However, the size of the region selected for the spectral extraction 
was not sufficiently small to resolve the Bondi radius. Therefore, the true temperature 
and density within the Bondi radius could be marginally different. This leads to the 
underestimation of the Bondi accretion rate which was estimated by using the relation 
$\dot{M}_{B} = \pi R_B^2 \rho c_s$. For the given ambient density and Bondi radius 
(\s 3.23 pc), the Bondi accretion rate was found to be $ 1.07\, \times\, 10^{-4}\,$ \Msun yr$^{-1}$ 
and provides a lower limit for a given back hole mass. Ratio of the required accretion 
rate ($\dot{M}_{acc}$) to the Bondi rate $\dot{M}_{B}$ suggest that the accretion rate 
required for carving X-ray cavities in this cluster is much higher, at least by a factor 
of one thousand than the Bondi rate.

%=================================================
\begin{figure}
\centering
\includegraphics[clip, trim=0.5cm 4.5cm 0.5cm 1cm, width=0.42\textwidth]{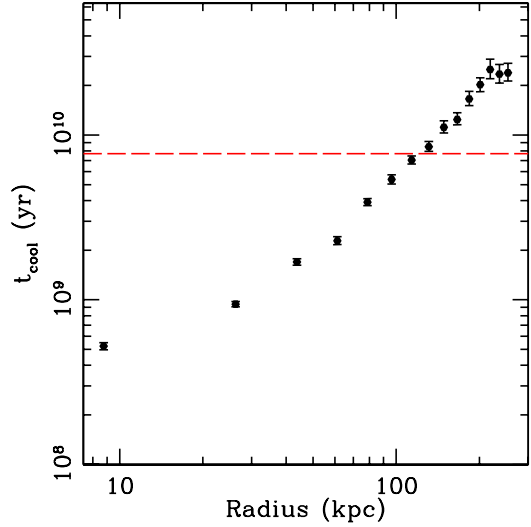}
\caption{Cooling time profile of the ICM in ZwCl~2701. The horizontal dashed line corresponds 
to the cooling time of 7.7 Gyr.}
\label{fig9}
\end{figure}
%=================================================

\subsection{Cooling versus Heating of the ICM}
In the absence of central heating, the ICM is supposed to radiatively cool and get deposited 
at the core of the cluster. The time required for cooling of the gas by radiating its enthalpy 
can be obtained by using the relation \citep{2009ApJS..182...12C}
\begin{equation}
t_{cool} = \frac{5}{2}\, \frac{nkT}{n_e n_H \Lambda(T,Z)}
\label{eq6}
\end{equation}
where, $n_e$ and $n_H$ are the electron and hydrogen densities, respectively, while 
$\Lambda(T,Z)$ represents the cooling function. We assume total number density $n= 2.3 n_H$ 
for the fully ionized plasma. Values of the $n_e$ and $n_H$ for ICM in ZwCl~2701 were 
derived by analyzing azimuthally averaged projected spectra from concentric annuli, as 
discussed earlier. The resultant cooling time profile of the ICM in ZwCl~2701 is shown 
in Figure~\ref{fig9} with the cooling time of the core of the cluster (within 5\arcsec) 
equal to \s $5 \times 10^8$ yr. The horizontal dashed line represents the ``cooling 
radius'' which is defined as the radius within which the gas has a cooling time less 
than 7.7 \tim 10$^9$ yr (\citealt{2004ApJ...607..800B}), also called the cosmological 
time. Following \cite{2012Natur.488..349M}, we estimate the cooling rate 
$dM/dt = 2 L_{X} \mu m_p/5kT$, $L_X$ being the 2-10 keV X-ray luminosity within cooling 
radius, $\mu$ the molecular weight and $kT$ the gas temperature, to be \s 196 \Msun 
yr$^{-1}$ within the cooling radius of $R_{cool}$ = 105 kpc\, (30\arcsec). 

%=================================================================
\begin{figure}
\centering
\includegraphics[trim={0 -0.8cm 0 0},clip=true, width=4.cm, height=3.9cm]{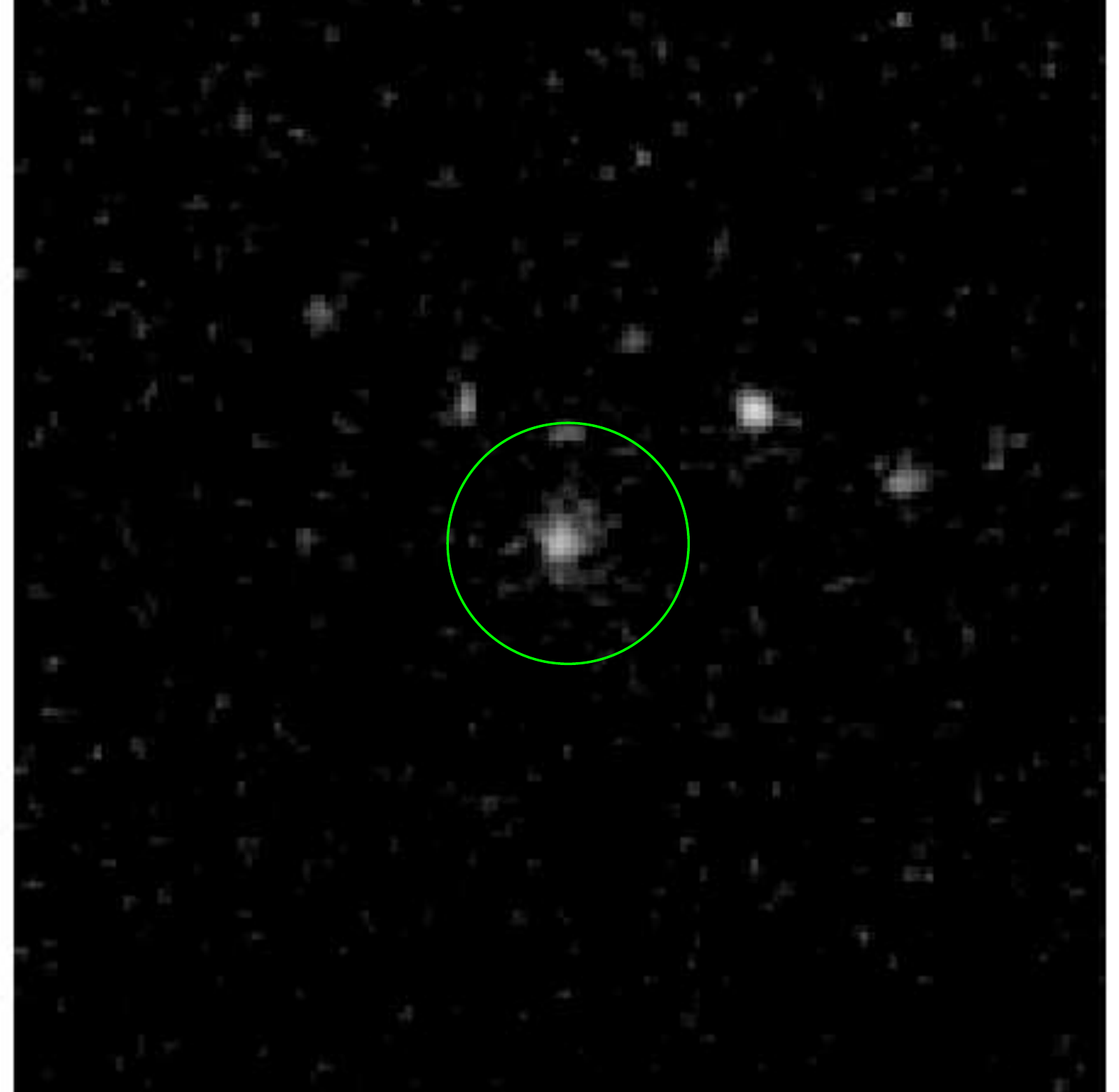} 
\includegraphics[width=4.1cm, height=4cm]{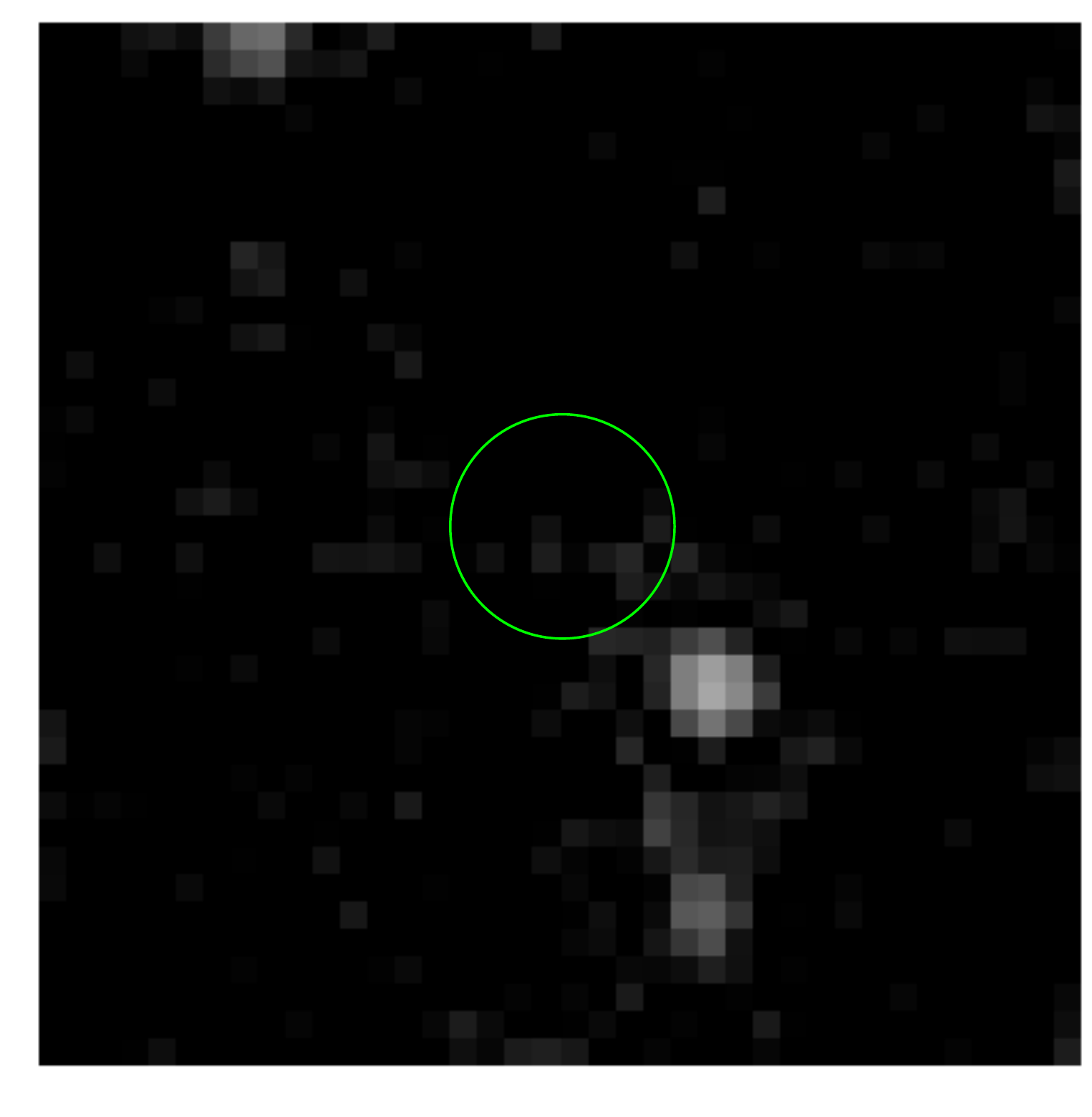} \\
\includegraphics[trim={0 5mm 0 5mm},clip=true, width=4.1cm, height=4cm]{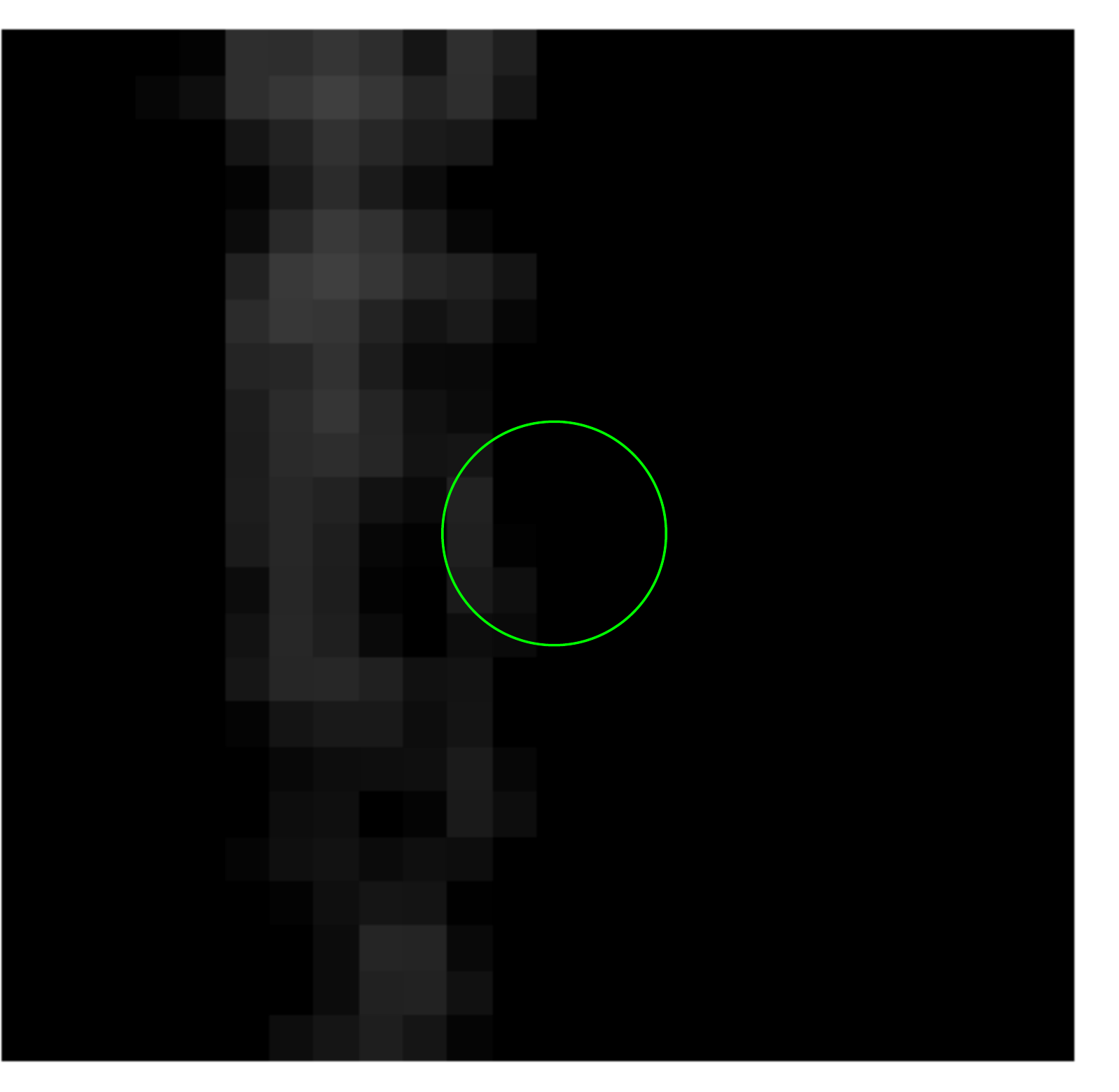}
\includegraphics[width=4cm, height=4cm]{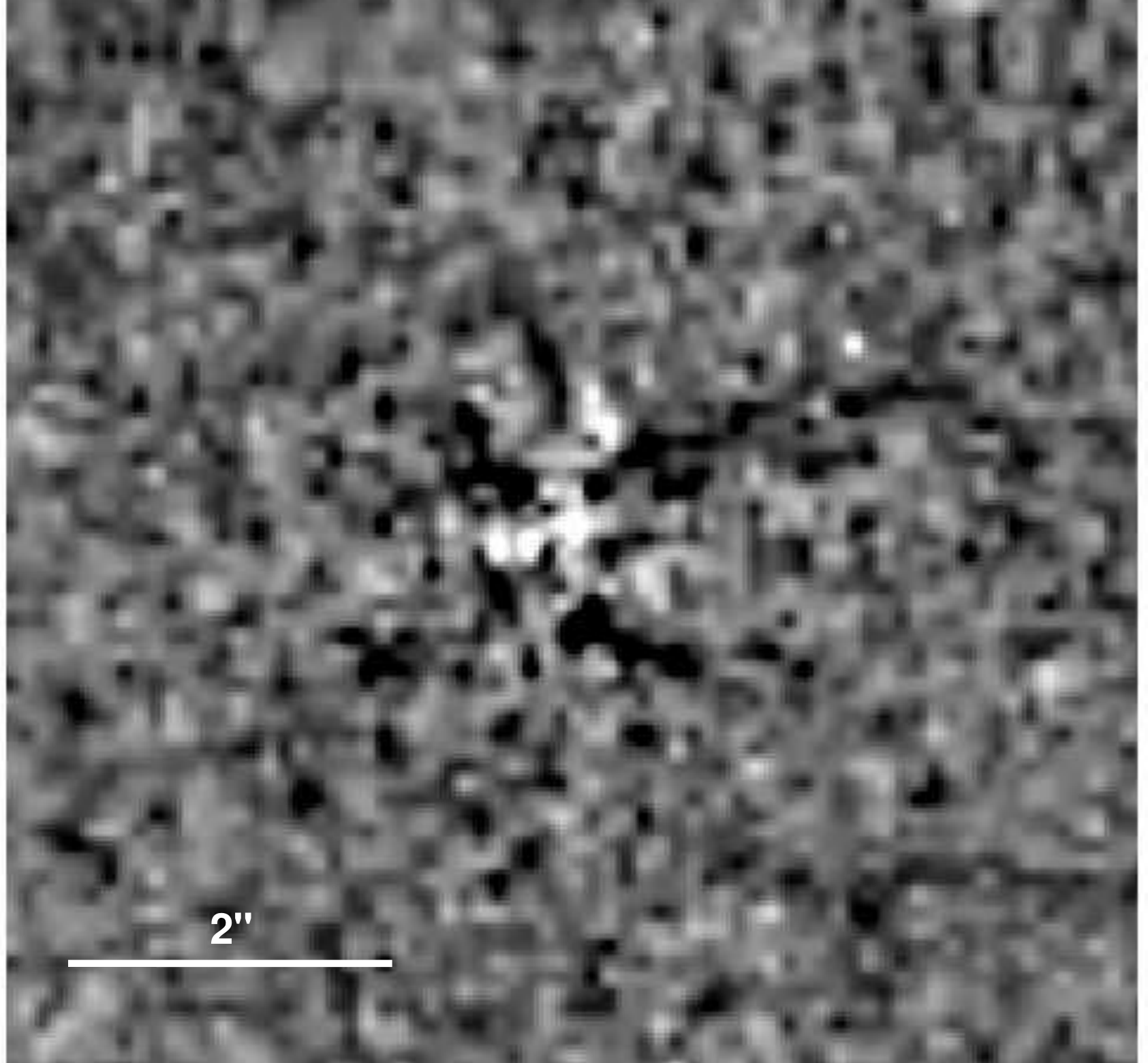}
\caption{\textit{(Top left):} Spitzer imagery at 8$\mu$m IR, \textit{(Top right):} 24$\mu$m MIR, 
\textit{(Bottom left):} 70$\mu$m FIR and \textit{(Bottom right):} \textit{HST} optical residual 
image generated from the optical \textit{HST} image peaked at center of the cluster 
(RA=9:52:49.168 DEC=+51:53:05.01). The green circles overlaid on the 8$\mu m$, 24$\mu m$ 
and 70$\mu m$ images indicate central 10\arcsec\, radius.} 
\label{fig10}
\end{figure}
%======================================================

Existence  of the cool molecular gas in the core of the cluster was investigated by 
employing the infrared (IR) observations of the ZwCl~2701. For this, we used \textit{Spitzer} 
archival IR and mid/far-infrared (MIR/FIR) data. \textit{Spitzer} IRAC 8$\mu m$ image (top left 
panel of Figure~\ref{fig10}) showed the presence of a bright point source within the radius of 
$\sim$4\arcsec, however, failed to exhibit the signature of the extended features. On the other 
hand, no emission was evident in the MIR 24$\mu m$ and FIR 70$\mu m$ images (top right and bottom 
left panel of Figure~\ref{fig10}). This confirms the marginal or no detection of the molecular 
gas in the core of ZwCl~2701. Further, residual image obtained after subtracting a smooth elliptical 
model of the \textit{HST} 606 $nm$ optical image \citep{2007A&A...461..103P,2012NewA...17..524V} 
could not substantiate the expected signatures of dust features (bottom right panel of 
Figure~\ref{fig10}). Marginal detection of dust and undetectable or very weak MIR/FIR emission 
from the core of this cluster indicate a very low rate of star formation. We made an attempt to 
quantify the star formation by using optical spectroscopic observations available in the archive 
of \textit{Sloan Digital Sky Survey} (SDSS)\footnote{\color{blue}{http://cas.sdss.org/dr7/en/tools/explore/obj.asp}}. 
The deblended $H_{\alpha}$ emission line flux ($F_{H\alpha} \sim 5.78 \times 10^{-16}\,$ \flux) 
yielded the star formation rate (SFR) of \s 0.60 \Msun yr$^{-1}$ (\citealt{1998ARA&A..36..189K}) 
which is very low compared to the expected one (\s 196 \Msun yr$^{-1}$) and hence confirms the 
cooling-flow problem in this cluster. A correlation between the BCG colour and separation
between the BCG center and X-ray peak is being discussed on the context of SFR (see Figure~11
of \citealt{2008ApJ...687..899R}). However, similar kind of investigation can not be done in the 
case of ZwCl~2701 due to the lack of UV data.

%=========================================
\begin{figure*}
\includegraphics[clip, trim=0.5cm 4.5cm 0.5cm 1cm, width=75mm,height=75mm]{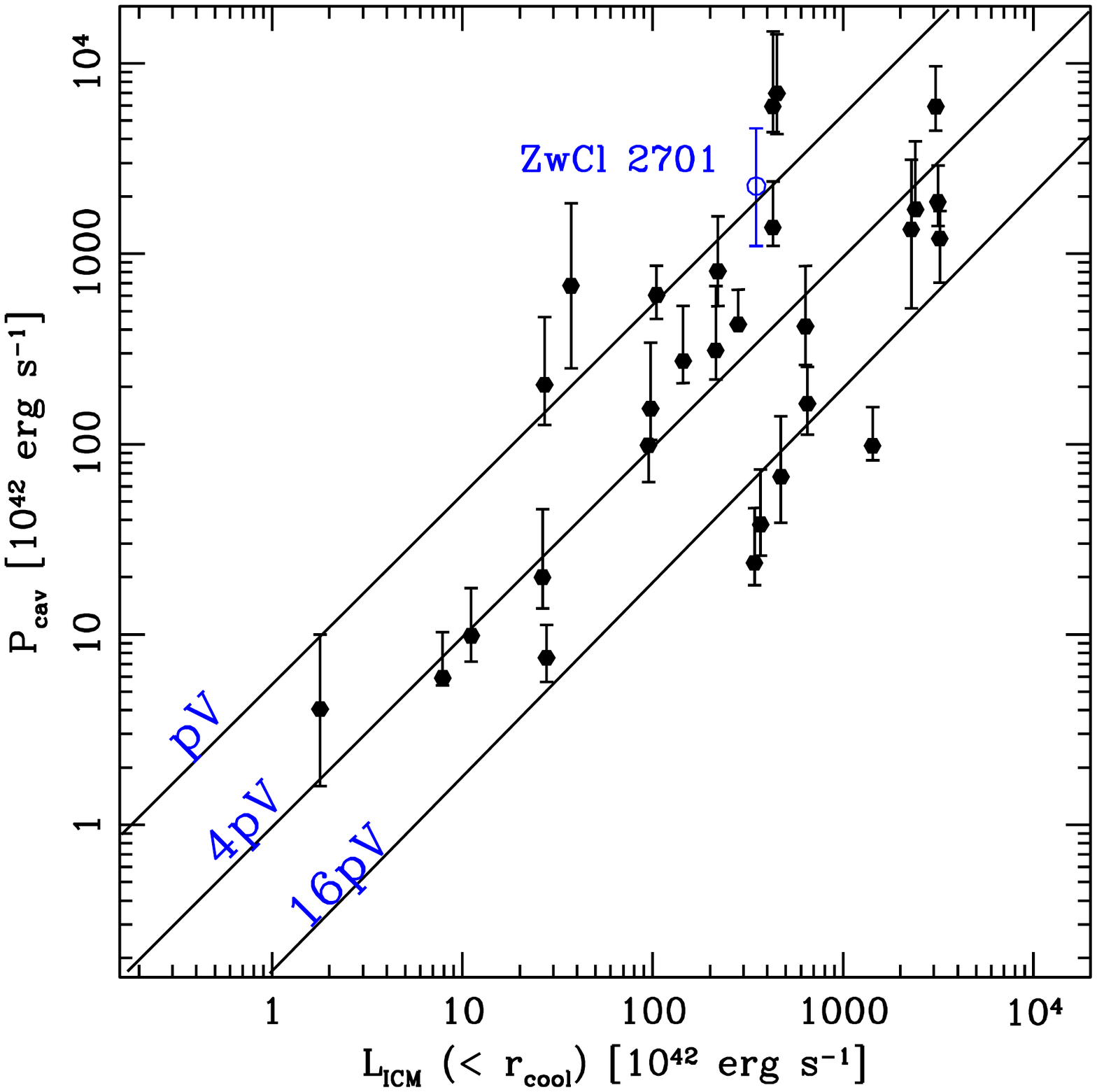}
\includegraphics[width=70mm,height=70mm]{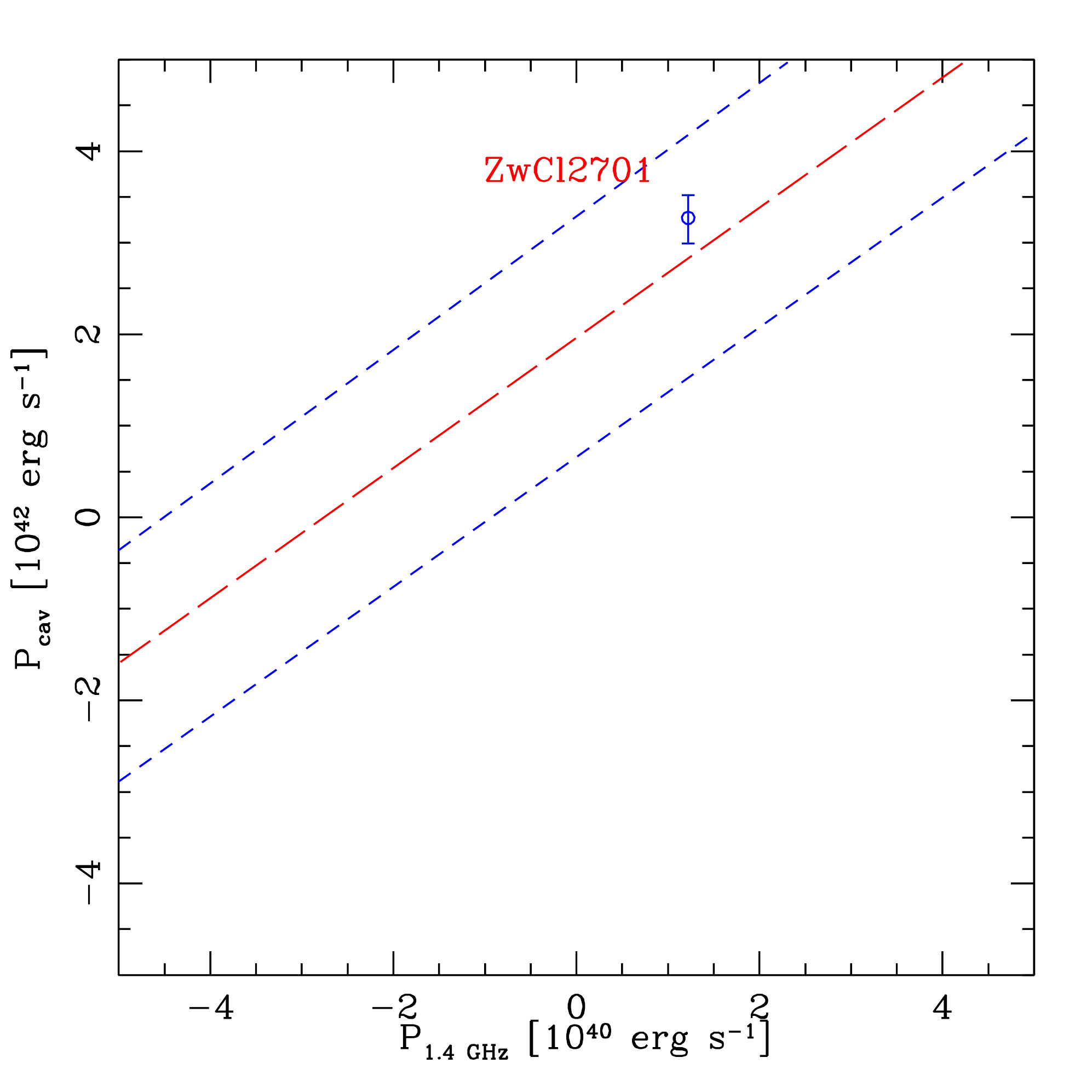}
\caption{\textit{Left panel:} Cavity heating power ($4pV$) against the total X-ray luminosity 
(within cooling radius). The diagonal solid lines denote cases where heating power equals to the
cooling power ($P_{cav} = L_{ICM}$) for $pV$, $4pV$ and $16pV$ enthalpy levels of the cavities. 
Filled circles indicate the cluster sample of \citet{2010ApJ...720.1066C}. The open circle in
the figure indicates the value of total cavity power of ZwCl~2701 cluster (present work).
\textit{Right panel:} Comparison between the cavity power $P_{cav}$ against 1.4 GHz radio 
power ($P_{1.4GHz}$) for the sample studied by \citet{2010ApJ...720.1066C}. The long dashed 
line represents the best fit relation for their sample of giant ellipticals (gEs) with the upper 
and lower limits shown by the short dashed lines. The estimated value for ZwCl~2701 cluster 
is marked in the figure.}
\label{fig11}
\end{figure*}
%=========================================

Evidence regarding the intermittent heating and offsetting of cooling in the core of the cluster 
was confirmed by the observed floor in the ``entropy profile'' of the ICM (\citealt{2000MNRAS.315..689L}). 
The entropy profile of the ICM in the environment of ZwCl~2701 was derived by performing 
spectral analysis of X-ray emission and using the thermodynamical parameters as derived above. 
The resultant entropy profile of the ICM is shown in Figure~\ref{fig4} \textit{(bottom panel)}. 
We tried to fit this profile by employing two different models, (i) a simple power law 
($S(r) = S_{100}\, (r/100kpc)^{\alpha_2}$) and (ii) a power law plus a constant term i.e., 
$S(r) = S_0 + S_{100}\, (r/100kpc)^{\alpha_1}$ (\citealt{2009ApJS..182...12C}) where $S_0$ 
represents the excess entropy value above the best fitted power law at larger radii and is 
known as the entropy floor. Power law plus a constant $S_0$ model resulted in the best fit 
(solid line in bottom panel of Figure~\ref{fig4}) of the entropy profile with the parameters 
$ S_0 = 19 \pm 1.38\,$ keV\,cm$^{2}$, $\alpha_1 = 1.41 \pm 0.05$ and $ S_{100} = 121.10 \pm 
3.56\,$ keV\,cm$^{2}$, and are consistent with those obtained by \cite{2009ApJS..182...12C} 
for a larger sample of cool core clusters. Similar results were also reported by 
\cite{2015Ap&SS.359...61S} for Abell 2390. The excess entropy $ S_0 = 23 \pm 1.38\,$ keV\,cm$^{2}$ 
in the core of this cluster confirms that some intermittent heating is operative in the core of 
ZwCl~2701.

The observed X-ray cavities, the floor in the entropy profile and the diffuse radio emission 
tracing the regions of X-ray cavities point towards the ongoing mechanical-mode feedback in 
the core of this cluster. To confirm whether the AGN feedback is powerful enough to quench 
cooling of the ICM in this cluster, we compare the cavity power ($P_{cavity}$) with the total 
radiative luminosity within the cooling radius ($L_{cool}$) as suggested by 
\citet{2006ApJ...652..216R} and is shown in Figure~\ref{fig11} \textit{(left panel)}. 
Filled circles represent the data points from the sample of \cite{2006ApJ...652..216R}. 
Position of the ZwCl~2701 is shown by the open triangle and occupies a position little 
above the 4$pV$ line. Solid diagonal lines in this plot represent the equivalence between 
the two ($P_{cav}$ = $L_{cool}$) at $pV$, 4$pV$ and 16$pV$ of the total enthalpy. For the 
case of ZwCl~2701, we estimated the cooling luminosity $L_{cool} = 3.5 \times\, 10^{44}\,$ \lum, 
while the total cavity power $P_{cav}$ \s $2.27 \times\, 10^{45}\,$ \lum, which means AGN 
feedback from the core offsets the cooling. Further confirmation of the balance between 
the two was provided by the comparison between the 1.4\,GHz radio luminosity and the cavity 
power. For this, we plot the measured value of GMRT 1.4 GHz total radio power of the source 
associated with ZwCl~2701 ($\sim$ 2.20 $\times$ 10$^{41}$ \lum) against the mechanical power 
of the X-ray cavities and is shown in Figure~\ref{fig11}\textit{(right panel)}. The long 
dashed line in this figure represents the best fit relation of \cite{2010ApJ...720.1066C} 
for a sample of giant ellipticals (gEs) with the upper and lower limits shown by the short 
dashed lines. In this plot, ZwCl~2701 occupies position close to the best-fit line obtained 
by \cite{2010ApJ...720.1066C}, suggesting that the radio source hosted by this system is 
powerful enough to meet the balance. The ratio between the two is $\sim 10^{-4}$, 
and is comparable to the measurements of cool core clusters (\citealt{2014arXiv1412.5664G,2008ApJ...686..859B}).

\section{Conclusions}
In this paper we presented a detailed multi-wavelength study of the cool core cluster 
ZwCl~2701. The study is based on the analysis of 123\,ks \textit{Chandra} and 8\,hrs 
\textit{GMRT}, archival optical (HST) and IR (Spitzer) data sets. The primary findings 
from this study are summarized as:
\begin{enumerate}
\item The morphological analysis of \textit{Chandra} image confirmed the presence of 
surface brightness depressions (cavities) in the central region of the cluster ZwCl~2701. 
\item 1.4 GHz radio map derived from the \textit{GMRT} data detected large scale radio 
emission that filled the X-ray cavities in the central region of the cluster. 
\item Comparison of the average cavity power (\s $2.27 \times\, 10^{45}\,$ \lum) with 
the X-ray luminosity of the gas within the cooling radius (\s 105 kpc) $L_{cool} \sim 
3.5 \times\, 10^{44}\,$ \lum suggested that the mechanical power of the AGN outburst 
is large enough to balance the radiative loss in the system.
\item Ratio of the 1.4\,GHz radio luminosity and the total power contained within the 
cavities is found to be $\sim 10^{-4}$ which is comparable to that of other cluster galaxies.
\item The back hole accretion rate in the brightest galaxy in ZwCl~2701 cluster was estimated to be
\s 0.11 \Msun yr$^{-1}$ which is well below the Eddington limit (\s 12 \Msun yr$^{-1}$) but above 
the Bondi accretion rate $1.07 \times 10^{-4}$ \Msun yr$^{-1}$. 
\end{enumerate}

\section*{Acknowledgements} 
We sincerely thank the referee for his/her valuable comments and suggestions which improved the paper.
The authors thank C. H. Ishwara-Chandra, Ruta Kale and Dharam V. Lal of NCRA, TIFR, India for their 
help in GMRT data analysis and inputs on the manuscript. MKP and SSS acknowledge the use of computing 
and library facilities of IUCAA, Pune, India and use of the High performance computing facility 
procured under the DST, New Delhi's FIST scheme (F No. SR/FST/PS-145/2009). This work has made 
use of data from the Giant Metrewave Radio Telescope (\textit{GMRT}), \textit{Chandra} data 
archive, \textit{SDSS} (Sloan Digital Sky Survey), archival data obtained with the \textit{Spitzer} 
Space Telescope, archival data from \textit{HST}, Extragalactic Database (NED) and and software 
provided by the Chandra X-ray Centre (CXC). SSS acknowledges financial support from the Ministry 
of Minority Affairs, Govt. of India, under the Minority Fellowship Program (Award No. 
F-17.1/2010/MANF-BUD-MAH-2111/SA-III/Website). 
%%%%%%%%%%%%%%%%%%%%%%%%%%%%%%%%%%%%%%%%%%%
%----------------- Bibliography and bibfile 
%------------------------------------------
\def\aj{AJ}%
\def\actaa{Acta Astron.}%
\def\araa{ARA\&A}%
\def\apj{ApJ}%
\def\apjl{ApJ}%
\def\apjs{ApJS}%
\def\ao{Appl.~Opt.}%
\def\apss{Ap\&SS}%
\def\aap{A\&A}%
\def\aapr{A\&A~Rev.}%
\def\aaps{A\&AS}%
\def\azh{AZh}%
\def\baas{BAAS}%
\def\bac{Bull. astr. Inst. Czechosl.}%
\def\caa{Chinese Astron. Astrophys.}%
\def\cjaa{Chinese J. Astron. Astrophys.}%
\def\icarus{Icarus}%
\def\jcap{J. Cosmology Astropart. Phys.}%
\def\jrasc{JRASC}%
\def\mnras{MNRAS}%
\def\memras{MmRAS}%
\def\na{New A}%
\def\nar{New A Rev.}%
\def\pasa{PASA}%
\def\pra{Phys.~Rev.~A}%
\def\prb{Phys.~Rev.~B}%
\def\prc{Phys.~Rev.~C}%
\def\prd{Phys.~Rev.~D}%
\def\pre{Phys.~Rev.~E}%
\def\prl{Phys.~Rev.~Lett.}%
\def\pasp{PASP}%
\def\pasj{PASJ}%
\def\qjras{QJRAS}%
\def\rmxaa{Rev. Mexicana Astron. Astrofis.}%
\def\skytel{S\&T}%
\def\solphys{Sol.~Phys.}%
\def\sovast{Soviet~Ast.}%
\def\ssr{Space~Sci.~Rev.}%
\def\zap{ZAp}%
\def\nat{Nature}%
\def\iaucirc{IAU~Circ.}%
\def\aplett{Astrophys.~Lett.}%
\def\apspr{Astrophys.~Space~Phys.~Res.}%
\def\bain{Bull.~Astron.~Inst.~Netherlands}%
\def\fcp{Fund.~Cosmic~Phys.}%
\def\gca{Geochim.~Cosmochim.~Acta}%
\def\grl{Geophys.~Res.~Lett.}%
\def\jcp{J.~Chem.~Phys.}%
\def\jgr{J.~Geophys.~Res.}%
\def\jqsrt{J.~Quant.~Spec.~Radiat.~Transf.}%
\def\memsai{Mem.~Soc.~Astron.~Italiana}%
\def\nphysa{Nucl.~Phys.~A}%
\def\physrep{Phys.~Rep.}%
\def\physscr{Phys.~Scr}%
\def\planss{Planet.~Space~Sci.}%
\def\procspie{Proc.~SPIE}%
\let\astap=\aap
\let\apjlett=\apjl
\let\apjsupp=\apjs
\let\applopt=\ao
\bibliographystyle{mn2e}
\bibliography{mybib}
\end{document}